\documentclass[10pt,journal,compsoc]{IEEEtran}

\usepackage{cite}
\usepackage{amsmath,amssymb,amsfonts}
\usepackage{textcomp}
\usepackage{xcolor} 
\usepackage{graphicx}
\usepackage{float}
\usepackage{subfig}
\usepackage{hyperref}
\usepackage{soul}
\usepackage{url}

\usepackage[c]{esvect}

\usepackage [ruled,linesnumbered] {algorithm2e} % \usepackage{algorithm}
\usepackage[noend]{algorithmic}
\usepackage{makecell}
\usepackage{cleveref}
\crefformat{section}{\S#2#1#3}
\crefformat{subsection}{\S#2#1#3}
\crefformat{subsubsection}{\S#2#1#3} 

\usepackage{pgfplots, pgfplotstable}
\pgfplotsset{compat=1.8}
\usepgfplotslibrary{statistics}

\usepackage{tikz}
\usetikzlibrary{chains}
\usetikzlibrary{arrows}
\usetikzlibrary{positioning}
\usetikzlibrary{shapes.geometric}
\usetikzlibrary{calc}
\usetikzlibrary{chains}
\usetikzlibrary{arrows}
\usetikzlibrary{positioning}
\usetikzlibrary{shapes.geometric}
\usetikzlibrary{calc}
\usetikzlibrary{automata}
\usetikzlibrary{scopes}
\usetikzlibrary{backgrounds}
\usetikzlibrary{patterns}
\usetikzlibrary{decorations.pathreplacing}
\usetikzlibrary{arrows, decorations.markings}
\tikzset{
	-|-/.style={
		to path={
			(\tikztostart) -| ($(\tikztostart)!#1!(\tikztotarget)$) |- (\tikztotarget)
			\tikztonodes
		}
	},
	-|-/.default=0.5,
	|-|/.style={
		to path={
			(\tikztostart) |- ($(\tikztostart)!#1!(\tikztotarget)$) -| (\tikztotarget)
			\tikztonodes
		}
	},
	|-|/.default=0.5,
	-|-|/.style 2 args={
		to path={
			(\tikztostart) -| ($(\tikztostart)!#1!(\tikztotarget)$) |- ($(\tikztostart)!#2!(\tikztotarget)$) -| (\tikztotarget)
			\tikztonodes
		}
	},
	-|-|/.default=0.5,
}

\tikzstyle{vecArrow} = [thick, decoration={markings,mark=at position
	1 with {\arrow[semithick]{open triangle 60}}},
double distance=1.4pt, shorten >= 5.5pt,
preaction = {decorate},
postaction = {draw,line width=1.4pt, white,shorten >= 4.5pt}]
\tikzstyle{innerWhite} = [semithick, white,line width=1.4pt, shorten >= 4.5pt]

\tikzset{
	invisible/.style={opacity=0},
	visible on/.style={alt={#1{}{invisible}}},
	alt/.code args={<#1>#2#3}{%
		\alt<#1>{\pgfkeysalso{#2}}{\pgfkeysalso{#3}} % \pgfkeysalso doesn't change the path
	},
}

\makeatletter
\pgfplotsset{
    boxplot prepared from table/.code={
        \def\tikz@plot@handler{\pgfplotsplothandlerboxplotprepared}%
        \pgfplotsset{
            /pgfplots/boxplot prepared from table/.cd,
            #1,
        }
    },
    /pgfplots/boxplot prepared from table/.cd,
        table/.code={\pgfplotstablecopy{#1}\to\boxplot@datatable},
        row/.initial=0,
        make style readable from table/.style={
            #1/.code={
                \pgfplotstablegetelem{\pgfkeysvalueof{/pgfplots/boxplot prepared from table/row}}{##1}\of\boxplot@datatable
                \pgfplotsset{boxplot/#1/.expand once={\pgfplotsretval}}
            }
        },
        make style readable from table=lower whisker,
        make style readable from table=upper whisker,
        make style readable from table=lower quartile,
        make style readable from table=upper quartile,
        make style readable from table=median,
        make style readable from table=lower notch,
        make style readable from table=upper notch,
        make style readable from table=draw position,
        make style readable from table=draw,
}
\makeatother

\newcommand{\SystemName}{{P4SGD}\xspace}

\newcommand{\allnotes}[1]{}
\renewcommand{\allnotes}[1]{#1} % Comment to turn off notes
\newcommand{\hhj}[1]{\textcolor{blue}{#1}}%hhj: 

\begin{document}
\title{\SystemName: Programmable Switch Enhanced Model-Parallel Training on Generalized Linear Models on Distributed FPGAs}

\author{Hongjing Huang, Yingtao Li, Jie Sun, Xueying Zhu, Jie Zhang, Liang Luo, Jialin Li, Zeke Wang$^*$ %~\IEEEmembership{Member,~IEEE}% <-this % stops a space

\IEEEcompsocitemizethanks{\IEEEcompsocthanksitem Hongjing Huang, Yingtao Li, Jie Sun, Xueying Zhu, Jie Zhang, Zeke Wang are with Collaborative Innovation Center of Artificial Intelligence, Zhejiang University, China. %\protect\\
E-mail: {huang\_hj, Li\_Yingtao, jiesun, zhuxueying, carlzhang4, wangzeke}@zju.edu.cn
\IEEEcompsocthanksitem Liang Luo is with University of Washington. E-mail: liangluo@cs.washington.edu % <-this % stops an unwanted space
\IEEEcompsocthanksitem Jialin Li is with National University of Singapore, Singapore.\protect\\
E-mail: lijl@comp.nus.edu.sg % <-this % stops an unwanted space

\IEEEcompsocthanksitem Jie Sun is also with Alibaba group, China. E-mail: sunjie.sun@alibaba-inc.com} % <-this % stops an unwanted space

\thanks{*: Corresponding author}}


\markboth{IEEE Transactions on Parallel and Distributed Systems}%
{Shell \MakeLowercase{\textit{et al.}}: Bare Demo of IEEEtran.cls for Computer Society Journals}


\IEEEtitleabstractindextext{%
\begin{abstract}
%Distributed machine learning (ML) systems are becoming increasingly prominent in recent years. The linear model is one of the most popular models for machine learning systems. However, the current linear model ML systems usually incur problems of the large batch size and huge communication overhead, which leads to slowly training convergence. To solve these problems, we are the first to propose that the P4 switch is used for the data aggregation of the model parallel distributed ML system. On this basis, we present a novel heterogeneous distributed ML system, which partitions training data and models by columns. It performs ML computation and network communication on the FPGA and aggregates the data of multiple workers through the P4 switch to reduce the time of data communication. We implemented it on Xilinx U280 and P4 switch. Experimental results show up to xxx faster than distributed linear model ML system which implemented on GPUs.

Generalized linear models (GLMs) are a widely utilized family of machine learning models in real-world applications. As data size increases, it is essential to perform efficient distributed training for these models. However, existing systems for distributed training have a high cost for communication and often use large batch sizes to balance computation and communication, which negatively affects convergence. 
% Generalized linear models (GLMs) are a family of the most widely used machine learning models in the real world. As the data size grows, efficient training of such models must be done in a distributed manner. However, existing distributed training systems suffer from large communication cost and often resort to large batch size to improve compute to communication ratio, which in turn hurts convergence. 
Therefore, we argue for an efficient distributed GLM training system that strives to achieve linear scalability, while keeping batch size reasonably low.
As a start, we propose \SystemName, a distributed heterogeneous training system that efficiently trains GLMs through model parallelism between distributed FPGAs and through forward-communication-backward pipeline parallelism within an FPGA. Moreover, we propose a light-weight, latency-centric in-switch aggregation protocol to minimize the latency of the AllReduce operation between distributed FPGAs, powered by a programmable switch. As such, to our knowledge, \SystemName{} is the first solution that achieves almost linear scalability between distributed accelerators through model parallelism. 
%a novel system \SystemName, that partitions training data and models by column, and leverages P4 switches and FPGAs for aggregation and computation to accelerate distributed linear model training in model parallelism. 
We implement \SystemName{} on eight Xilinx U280 FPGAs and a Tofino P4 switch. Our experiments show \SystemName{} converges up to 6.5X faster than the state-of-the-art GPU counterpart.  
\end{abstract}
}

\maketitle

\IEEEdisplaynontitleabstractindextext

% \fi
\IEEEpeerreviewmaketitle

% \IEEEraisesectionheading{\section{Introduction}\label{sec:introduction}}

\vspace{-1.5ex}
\section{Introduction}
\label{intro}
\vspace{-0.5ex}

\begin{figure}[t]
	\centering
	\subfloat[Data parallel]{\includegraphics[width=1.5in]{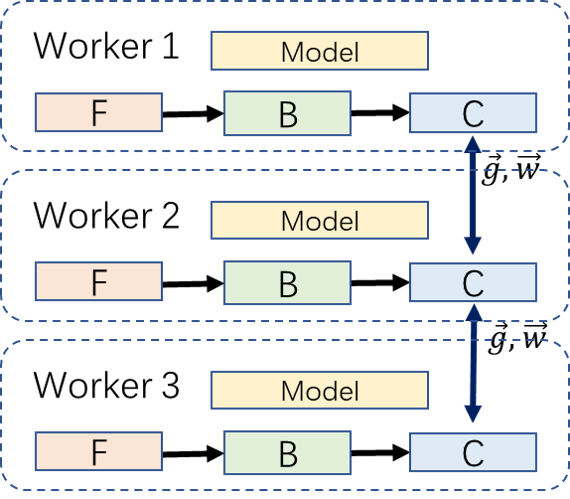} \label{data_parallel_f}} % \caption{}
	\subfloat[Model parallel]{\includegraphics[width=1.7in]{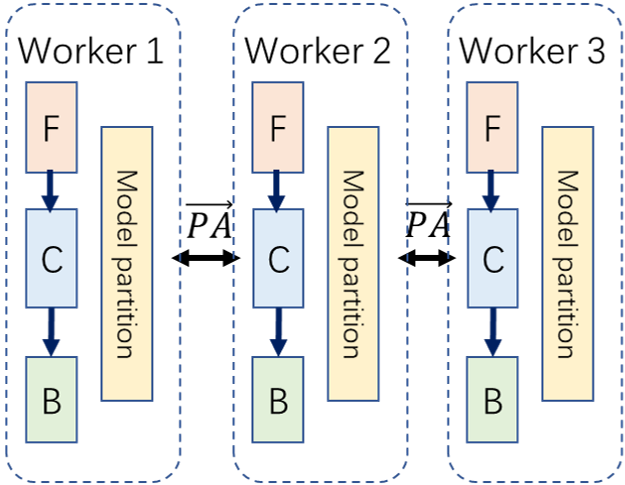} \label{model_parallel_f}}
	\vspace{-1ex}	
	\caption{Comparison of data-parallel and model-parallel training, F: forward propagation, B: backward propagation, C: communication, $\vec{g}$: gradient, $\vec{w}$: model, $\vec{PA}$: partial activations. Data (or model) parallelism needs to collectively communicate the whole gradient (or a mini-batch of activations) per iteration.} 
	\vspace{-3ex}
	\label{dp_mp_intro} 
\end{figure}

\IEEEPARstart{M}{achine} learning (ML) is a popular approach to accurately predict outcomes without needing to be explicitly programmed to do so. 
While most of the previous work focuses on accelerating deep neural network model training, generalized linear models (GLMs), such as linear regression, classification, and support vector machine~\cite{rendle2010factorization}, remain one of the most widely used models in the real world~\cite{elgohary2016compressed,liu2015efficient}. 
% \marginpar{R2.O5-1}\hhj{\st{In recent years, it has played a huge role in many aspects. The linear model machine learning system is a very useful part of machine learning. }}
 Because the number of training samples has grown from the previously tens of thousands to tens of millions today~\cite{zheng2019real,pop2016design} and a single machine typically does not feature enough computing power and memory capacity to allow efficient GLM training on a large number of samples. Thus, people resort to distributed GLM training. 
%  Even though GLM, literally similar to a fully-connected layer, has a low computation to communication ratio, training GLMs still raises a severe performance issue.
\emph{Data-parallel} and \emph{model-parallel} are two of the prevailing parallel paradigms for distributed training.
Both approaches consist of three stages: \emph{forward propagation}, \emph{backward propagation}, and \emph{communication}, as shown in Figure~\ref{dp_mp_intro}. 

%\liang{we probably don't need the ''which are adopted by most distributed ML frameworks...`` if we need more space.}

%In the following, we discuss the hardware implementation difference between data-parallel and model-parallel training on GLMs. Either approach consists of three stages: \emph{forward propagation}, \emph{backward propagation}, and \emph{communication}, as shown in Figure~\ref{dp_mp_intro}. Each stage has its own hardware resources to implement, such that inter-stage parallelism is achieved whenever no dependency exists. %In addition, The computing power of a single machine is limited, to speed up the training systems. There is wide research and engineering efforts on distributed training systems. Distributed ML can be divided into two types, data parallel and model parallel. 

%To make an informed choice of parallelism paradigm, 
%We compare data and model parallelism in the context of training GLMs, in terms of their compute and communication characteristics, as numerous work~\cite{sapio2019scaling, luo2018parameter} has pointed out that communication has become the major bottleneck of distributed training. Both paradigms require careful engineering to fully overlap communication and computation when no dependency exists to achieve good performance.

% (Algorithm Optimization)
% data parallel vs. model parallel
% why model parallel?
% XXXXXX
% XXXXXX

% (System Optimization)
% CPU, GPU, ...

% Spark, Angel, SINGA, 
% Tensorflow (layer-wise).

% in-layer partition, inter-layer partition.

% why FPGA?
% model parallel is sensitive to latency.

% (Technical Contribution)

%~\liang{check caption: for data parallel, the communication cost is independent of batch size because the gradient is with respect to the model.}

\noindent{\bf Data Parallelism. }Data parallelism horizontally partitions the input dataset across workers, i.e., accelerators. Within each iteration, each worker uses its local copy of the entire model ($\vv{w}$) to train on its subset of the dataset, as illustrated in Figure~\ref{data_parallel_f}. 
% We observe that forward and backward propagation stages do not have any dependency and then execute concurrently. 
At end of each iteration, gradients of the model ($\vv{g}$) are averaged across all workers through \texttt{AllReduce}~\cite{oden2014energy,lee2020flexreduce} or parameter servers~\cite{li2014scaling,ho2013more,cui2016geeps}.

\noindent{\textbf{Model Parallelism}}. Model parallelism vertically partitions both the model $\vv{w}$ and the input dataset, such that each worker trains its model subset on its dataset subset, as shown in Figure~\ref{model_parallel_f}. Model parallelism requires synchronizing ``partial activations ($\vv{PA}$)" between workers.

Both forward and backward passes in GLM training involve a low computation per weight ratio. Therefore, both parallelisms need to frequently synchronize the model (or partial activations), which significantly impacts performance when more accelerators, e.g., GPUs, are involved.
% \marginpar{R1.O1}\hhj{
When a model grows to millions of parameters, data parallelism requires transmitting millions of gradients or models through the network, resulting in significant communication overhead. In contrast, model parallelism avoids exchanging gradients and models and transmits only activations. This significantly reduces communication overhead and gives model parallelism higher potential than data parallelism when training a large GLM, such as a fully-connected layer, on multiple accelerators.
% Compared with data parallelism, model parallelism incurs less network traffic, so model parallelism has more potential when training a large GLM, i.e., a fully-connected layer, on multiple accelerators.
\footnote{It is coincident with the observation from Krizhevsky~\cite{dp_mp_mix_Krizhevsky14}, who proposes to train convolutional layers through data parallelism due to their high computation amount per weight, and to train fully-connected layers through model parallelism due to low computation amount per weight.} 
However, it still faces one severe issue.

% \marginpar{R1.O2}\hhj{
Vanilla model-parallel training, i.e. basic model-parallel training, as shown in Figure~\ref{model_parallel_f}, requires an AllReduce operation to gather partial activations between forward and backward propagation. This exchange of partial activations creates a dependency that makes it impossible to overlap the forward and backward propagation.
% Vanilla model-parallel training, i.e. basic model-parallel training, illustrated in Figure~\ref{model_parallel_f}, needs an {AllReduce} operation to collect partial activations between forward and backward propagation, and thus cannot overlap forward and backward propagation due to the dependency introduced by the exchange of partial activation.
Therefore, model parallelism is more sensitive to both throughput and latency~\cite{shoeybi2020megatronlm, jia2018data, terapipe_arxiv21, ZeRO}, different from data parallelism which is more sensitive to the aggregation throughput~\cite{sapio2019scaling, lao2021atp}. 
% \marginpar{R1.O3-1 R1.O3-2}\hhj{
It has become widely accepted that model parallelism can only achieve linear scalability between accelerators within a node~\cite{shoeybi2020megatronlm, jia2018data, terapipe_arxiv21, ZeRO} that are connected with high-bandwidth, low-latency NVLinks. 
Furthermore, due to limited network bandwidth and high network latency between nodes, strong scaling (linear speedup from using more accelerators to train a model under the same whole mini-batch size) with multiple accelerators in model parallelism remains challenging.
% It becomes a common wisdom that model parallelism only achieves linear scalability between accelerators within a node~\cite{shoeybi2020megatronlm, jia2018data, terapipe_arxiv21, ZeRO}, where accelerators are connected with high-bandwidth and low-latency NVLinks; while model parallelism cannot achieve almost strong scaling (linear speedup from using more accelerators to train a model under the same whole mini-batch size) between distributed accelerators, due to limited network bandwidth and high network latency between nodes.
In this paper, we ask: 
%The common wisdom is that model parallelism works efficiently between accelerators within a node, rather than between nodes,  

%Second, the overhead of communication, particularly a collective operation \texttt{AllReduce} per iteration, could significantly impede us from achieving excellent strong scaling (linear speedup from using more accelerators to train a model under the same whole mini-batch size). In this paper, we ask: %~\liang{allreduce is point to point}
%\vspace{0.5em}
\begin{center}
{\em Can we achieve strong scaling when training a large GLM through model parallelism between distributed accelerators?}
\end{center}
%\vspace{0.5em}
% \hhj{To solve this problem, we consider the use of FPGAs for GLM training. Compared to general processors, e.g. CPUs or GPUs, the main advantage of FPGAs is their real-time and flexibility. We can implement the network module on FPGAs instead of connecting additional network cards, which significantly reduces the delay. At the same time, FPGAs can achieve finer-grained pipeline parallelism than CPUs or GPUs. }
To answer this question, as a start, we propose \SystemName, an efficient model-parallel training system that enables strong scaling when training a GLM.\footnote{We will generalize \SystemName{} to Deep Learning (DL) model training in future work. We believe \SystemName{} has great potential on DL model training as well. } \SystemName{} consists of a server implemented on a programmable switch and multiple workers implemented on FPGAs. 
\SystemName{} has three key innovations to address the above issue: efficient model-parallel training between distributed accelerators ({\bf C1}), forward-communication-backward pipeline parallelism within an accelerator ({\bf C2}), and ultra-low-latency aggregation between a P4 switch and FPGAs ({\bf C3}). 

\noindent
{\bf C1: Efficient Model-parallel Training between Distributed Accelerators. } Model parallelism works only well between accelerators, e.g., GPUs, within a node, which features high-bandwidth and low-latency links, e.g., NVLink, to connect these accelerators~\cite{shoeybi2020megatronlm, jia2018data, terapipe_arxiv21}. The key to scalability lies in ultra-low AllReduce latency and the overlap between the communication stage and forward/backward stages. To our knowledge, we are the first to propose \SystemName{} to achieve linear scale-out training on a GLM model, e.g., fully-connected layer, through model parallelism between distributed accelerators, e.g., FPGAs, which are connected via Ethernet. 

%The common wisdom says that   
%In order to address the first issue, each FPGA-based worker implements forward/backward propagation and communication with their own hardware resources, and divides each mini-batch of samples into multiple smaller \emph{micro-batches}, which flow into three stages without suffering from any dependency between each other. As such, \SystemName{} can explore the overlap between forward/backward propagation and communication from different micro-batches.  

%\vspace{0.3em}
\noindent
{\bf C2: Forward-Communication-Backward. } Pipeline parallelism within an FPGA. %In order to address the first issue, 
\SystemName{} exploits a new parallelism dimension, i.e., \emph{forward-communication-backward pipeline parallelism}, to maximize compute efficiency. From a hardware perspective, \SystemName{} implements three distinct execution stages (i.e., forward propagation, communication, and backward propagation) with their own hardware resources, and thus organizes GLM training into three execution stages to allow pipelined execution between stages. From a software perspective, \SystemName{} divides each mini-batch of samples into multiple smaller \emph{micro-batches}, which flow into three execution stages without suffering from any dependency on each other. As such, \SystemName{} can explore the overlap between forward/backward propagation and communication from different micro-batches, and also minimize communication overhead from model-parallel training.  %\wzk{We exploit a new parallelism dimension, hardware level, for pipeline-parallel training of GLM. Essentially, \SystemName{} allows overlapping forward and backward propagation. }%minimize the negative effect of communication needed by model-parallel training between distributed FPGAs

%\vspace{0.3em}
\noindent
{\bf C3: Latency-centric In-switch Aggregation Protocol. }We design and implement a light-weight, fault-tolerant, \emph{latency-centric} in-switch aggregation mechanism that directly interplays between a P4 switch and distributed FPGAs to minimize the latency of \texttt{AllReduce} needed by each training iteration. Such an aggregation protocol needs a careful interplay between distributed FPGAs and a P4 switch to recover from potential packet loss while achieving ultra low, stable {AllReduce} latency due to pure hardware implementation. %To address the second issue, 
%However, the communication between P4 switch and FPGA could be unreliable, so we propose a light-weight, fault-tolerant, \emph{latency-centric} in-switch aggregation protocol, which needs the careful interplay between distributed FPGAs and a P4 switch, to recover from packet loss with the goal of minimal latency overheads. As such, \SystemName{} achieves extremely low \texttt{AllReduce} latency. % on a P4 Switch

We implement the workers of \SystemName{} on up to eight Xilinx U280 FPGA boards~\cite{u280} and the server on a Tofino P4 switch~\cite{Wedge100BF}. The experimental results show that 1) \SystemName{} under model parallelism achieves strong scaling between distributed FPGA-based accelerators; 
2) \SystemName{} converges faster than its corresponding data-parallel counterpart; %in-network aggregation can reduce the latency of the collective operation between distributed FPGAs to 1.2$\mu$s, which is, to our knowledge, the smallest between distributed computing nodes; 
and 3) \SystemName{} converges up to 9.3X faster than the state-of-the-art distributed training systems on distributed GPUs that are not able to achieve linear scale-out training mainly due to its high inter-GPU communication overhead.

\vspace{-1.5ex}
\section{Background: Paralleling SGD Hardware}
In this section, we briefly discuss the interesting properties of the hardware implementation of data-parallel and model-parallel training on GLMs. The stochastic gradient descent (SGD) hardware through either model or data parallelism consists of three stages: forward propagation, backward propagation, and communication. All three stages, which have their own hardware resources to implement, allow to explore more parallelism. Table~\ref{t_time_comparation} illustrates the comparison result between data parallelism (``DP") and model parallelism (``Vanilla MP"). 
%compare the typical SGD hardware implementation between data-parallel and model-parallel. 

\begin{table} [t]
\renewcommand\arraystretch{1.4}
\setlength\tabcolsep{3pt}
	\centering

	%\begin{spacing}{0.3}
		\begin{scriptsize}
		
	\vspace{-0.5ex}
	\caption{Data parallelism (DP) vs. model parallelism (MP). $D$: model dimension, $M$: number of workers, $S$: number of samples, $B$: mini-batch size, $BW$: aggregation bandwidth between workers, $T_{f\_D}$: forward propagation time of DP, $T_{f\_M}$: forward propagation time of MP, $T_l$: aggregation latency, $T_{b\_D}$: backward propagation time of DP, $T_{b\_M}$: backward propagation time of MP, $MB$: micro-batch size.} \vspace{-0.5ex}
	\label{t_time_comparation}
	\begin{tabular}{|c||c|c|c|c|}
		\hline
		{ }  &  \makecell{{\bf Model} \\ {\bf mem.}}  &  \makecell{{\bf Dataset} \\ {\bf mem.}}  &  \makecell{{\bf Network} \\ {\bf mem.}} &  {\bf Iteration time} \\
		\hline
		\hline
		{DP}  & $D$ & $\frac{S \times D}{M}$ & $D$ & $T_{f\_D} + \frac{T_{b\_D}}{B} + \frac{D}{BW} + T_l$\\ 
		\hline
		{Vanilla MP} &  $\frac{D}{M}$ &  $\frac{S \times D}{M}$ & $B$ & $T_{f\_M} + T_{b\_M} + \frac{B}{BW} + T_l$\\ 
		\hline
		{\SystemName{} MP} & $\frac{D}{M}$ &  $\frac{S \times D}{M}$ & $B$ & $\frac{MB}{B} * T_{f\_M} + T_{b\_M} + \frac{MB}{BW} + T_l$  \\ 
		\hline
	\end{tabular}
	\vspace{-1ex}

		\end{scriptsize}
	%\end{spacing}
\end{table}
%Due to the large dataset and model, people propose various distributed versions of SGD. All exiting distributed SGD can be divided into two type: data-parallel SGD and model-parallel SGD. We then introduce these two different distributed SGD strategies.

\subsection{Data-Parallel Training}%$\\$ 
Data parallelism horizontally partitions the input dataset across workers, as shown in Figure~\ref{data_parallel_f}. Each worker maintains a local copy of the model and trains on its own partition of the input dataset. During each iteration, each worker goes through three steps, as shown in Figure~\ref{dp_overlap}. First, it computes the dot product of the updated model and each sample in the current mini-batch ({forward propagation}). Second, it uses the dot products from the mini-batch to compute the gradient ({backward propagation}). Third, it synchronizes its gradient with other workers and computes the updated model via either collective primitive \texttt{AllReduce} or parameter server~\cite{li2014scaling,ho2013more,cui2016geeps}. The SGD hardware has custom hardware resources for both forward and backward passes, so after we perform a forward pass on a sample, the backward pass can be started immediately without waiting for the forward results of other samples in the same mini-batch. Therefore, forward and backward passes have no dependency between samples in the same mini-batch. %

\noindent{\bf Estimating Elapsed Time per Epoch. }Since the three stages have their own forward-communication-backward pipeline, they can execute concurrently when there is no dependence between them. In particular,
% \marginpar{R2.O2-1} 
each time the forward propagation computes the loss of a sample, the backward propagation of this sample can be started immediately, so these two stages allow concurrent execution, but still have a dependency on the third stage communication. Therefore, time $T_{it}$ per iteration is estimated to be forward propagation time of DP $T_{f\_D}$ plus backward propagation time of DP for a sample $\frac{T_{b\_D}}{B}$ plus communication time $T_c$, as shown in Equation~\ref{equ_dp}.
\begin{equation}
\label{equ_dp}
    T_{it} = T_{f\_D} + \frac{T_{b\_D}}{B} + T_c = T_{f\_D} + \frac{T_{b\_D}}{B} + \frac{D}{BW} + T_l, 
\end{equation}
where $D$ is the model dimension, $BW$ is the aggregation bandwidth between workers, and $T_l$ is the aggregation latency. 

\subsection{Model-Parallel Training}%$\\$
\begin{figure}[t]
	\centering
	\subfloat[Data-parallel training]{\includegraphics[width=3.4in]{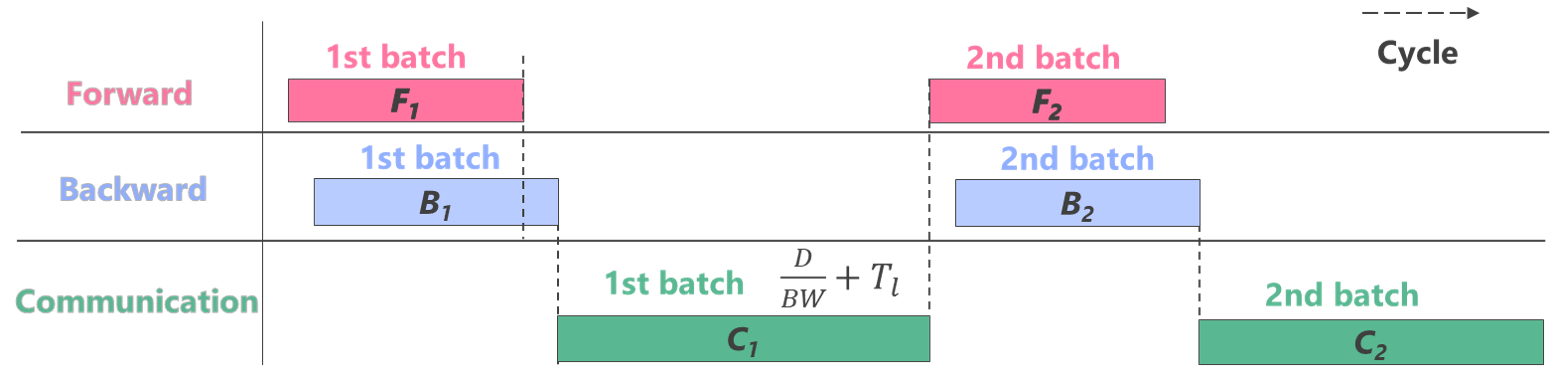} \label{dp_overlap}} %\vspace{-1ex}
    \hfill
    \vspace{-1ex}
    \subfloat[Vanilla mini-batch model-parallel training]{\includegraphics[width=3.4in]{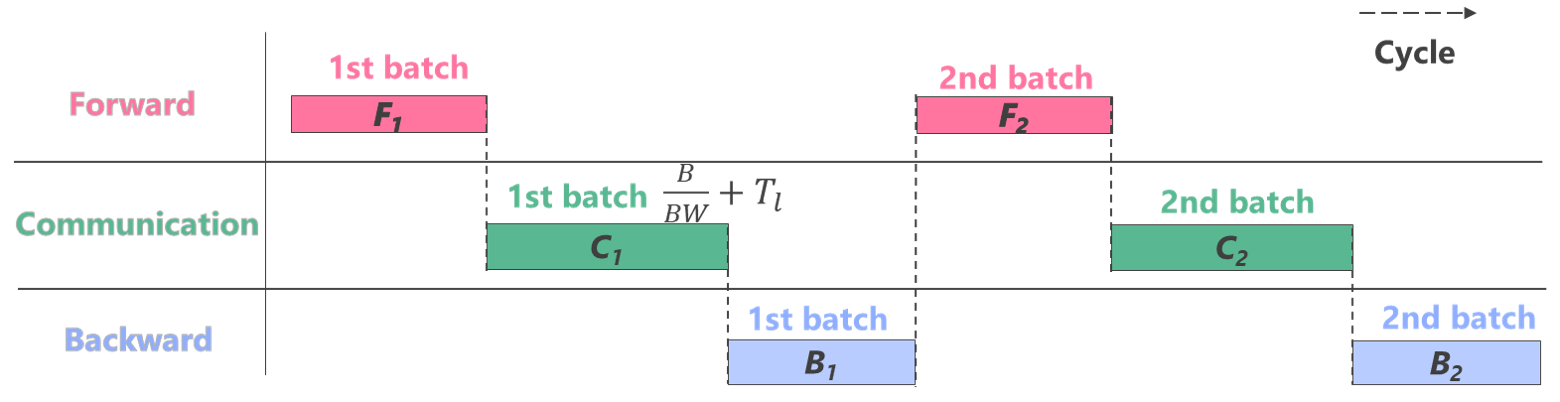} \label{mp_overlap}} %\vspace{-1ex}
    \hfill
	\subfloat[Micro-batch forward-communication-backward pipeline-parallel training within a worker]{\includegraphics[width=3.4in]{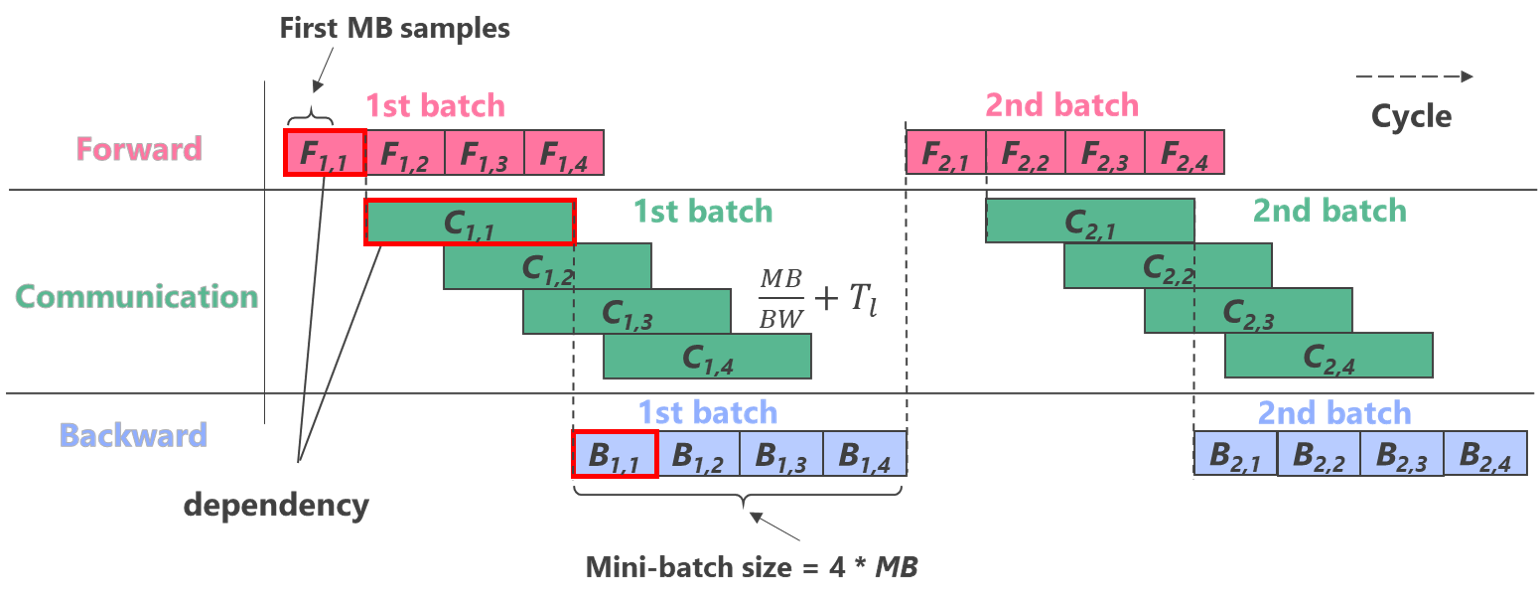} \label{mp_microbatch}}
	\vspace{-1ex}	
    `	\caption{Comparison of data and model parallelism for training GLMs implemented in hardware, motivating forward-communication-backward pipeline parallelism. $F_{i,j}$: forward propagation of the $j$-th micro-batch of the $i$-th mini-batch, $B_{i,j}$: backward propagation of the $j$-th micro-batch of the $i$-th mini-batch, $C$: communication, MB: micro-batch size, BW: network bandwidth, $T_l$: fixed latency of network } 
	\vspace{-3.5ex}
	\label{hbm_read} 
\end{figure}

\label{model_parallel_analyse}
Model parallelism (MP) vertically partitions the model across workers, as shown in Figure~\ref{model_parallel_f}. Each worker maintains a partition of the model and trains on its own partition of the input dataset. During each iteration, each worker goes through three steps, as shown in Figure~\ref{mp_overlap}. First, it computes the partial dot product of the partial model and each partial sample in the current mini-batch ({forward propagation}). Second, it synchronizes its partial dot products from the mini-batch with other workers and computes the full dot product via collective communication primitive \texttt{AllReduce}. Therefore, the amount of data to be exchanged is $B$, which is the batch size. Third, each worker uses the full dot products to compute its gradient portion ({backward propagation}). 

\noindent{\bf Estimating Elapsed Time per Epoch. }Since three stages have the dependency regarding the model, their executions are serialized. Therefore, $T_{it}$ is estimated to be forward propagation time of MP $T_{f\_M}$ plus backward propagation time of MP $T_{b\_M}$ plus communication time $T_{c}$, as shown in Equation~\ref{equ_mp}.
\begin{equation}
\label{equ_mp}
    T_{it} = T_{f\_M} + T_{b\_M} + T_c = T_{f\_M} + T_{b\_M} + \frac{B}{BW} + T_l 
\end{equation}
%where $B$ is the mini-batch size. 

\subsection{Data- vs. Model-Parallel Training} %$\\$Scale-out
We compare the scale-out potential between data-parallel training with model-parallel training on $M$ workers. From Equations~\ref{equ_dp} and~\ref{equ_mp}, intuitively, we observe that data parallelism allows the overlap between forward propagation and backward propagation, but needs to communicate the whole gradient during each iteration. Model parallelism only needs to communicate $B$ elements per iteration, but cannot overlap forward and backward propagation due to the dependency. In the following, we discuss their potential on scale-out training.

\noindent{\bf Potential of Data Parallelism on Scale-out Training. }Data parallelism spans a mini-batch of samples to $M$ workers, such that each worker only needs to work on $\frac{B}{M}$ samples, indicating that forward propagation time can be reduced by $M$ times. However, its communication time stays the same, because the amount of data to be exchanged stays the same. However, in practice, its communication time would increase. A larger $M$ could easily make data parallelism communication-bound, especially on GLMs that have a relatively low amount of computation per weight. Another drawback is that data parallelism fails to support a large model that exceeds the memory capacity of a worker. 

\noindent{\bf Potential of Model Parallelism on Scale-out Training. }Model parallelism spans the model and the dataset to $M$ workers, such that each worker only has a partition of the model and forward and backward propagation time can be reduced by $M$ times. Moreover, model parallelism only needs to exchange $B$ elements with other workers, incurring negligible overhead.\footnote{Even though the latency regarding $B$ elements is small, the basic network latency $T_l$ is still large.} Therefore, model parallelism can easily scale out to support $M$ workers for training. 
%In data-parallel SGD, the server stores the model, and each worker stores a horizontal partition of the dataset. In every iteration, the process is as follows. First, the workers pull the model from the server. Second, each worker performs two stages: forward propagation and backward propagation. Third, the workers push the gradient to the server. Forth, the server update the model.  

%In model-parallel SGD, the server aggregates the data, and each worker stores a vertical partition of the dataset as well as the model. In every iteration, the process is as follows. First, each worker compute the $partial\ activations\ (PA)$ by its own dataset and model. Second, the workers push the $PA$ to the server. Third, the server sum up the $PA$ and get the $full\ activations\ (FA)$. Forth, the workers pull the $FA$ from the server. Fifth, each worker compute the gradient by the $FA$ and its own dataset. Sixth, each worker updates its model.

\vspace{-1.5ex}
\section{System Overview of \SystemName}
%\SystemName{} intends to leverage the main advantage of model-parallel training while minimizing its drawbacks. In the following, we highlight three design goals, followed by the concrete design of \SystemName{} to achieve these goals.

\subsection{Design Goals and Overall Architecture}
When designing \SystemName{}, we keep three goals in mind.

\noindent{\bf G1: Allowing Efficient Scale-out Training. } 
As the model size of GLM is ever increasing, it is not appropriate to train the model on a single worker that has limited memory and compute capacities. Therefore, it is natural to employ multiple workers to concurrently train the same model. However, it is challenging to achieve linear scalability, since GLM has low computation amount per weight, making communication overhead difficult to amortize on CPUs/GPUs.    

\noindent{\bf G2: Maximizing the Overlap between Forward and Backward Propagation Computation. }The SGD hardware through vanilla model parallelism allows no overlap between forward and backward propagation due to its inherent dependency, resulting in low utilization of computing resources. Therefore, maximizing the overlap between forward and backward propagation could significantly reduce computation time and then improve the efficiency of model-parallel training.% on GLMs.  

\noindent{\bf G3: Minimizing Communication Latency. }Even the network traffic needed by model-parallel training is $B$ elements per iteration, network latency $T_l$ could still impede linear scalability, especially when $M$ is large. This is because $M$ workers lead to $M$ times less computation time per iteration, indicating a larger proportion of time spent on communication. Therefore, minimizing the latency of communication primitive AllReduce among workers could significantly benefit scale-out training.   

% \begin{figure} 
% 	\centering
% 	\includegraphics[width=3.4in]{figure/system_architecture.png} 
% 	% \caption{}
% 	\vspace{-1ex}		
% 	\caption{System architecture} 
% 	%\vspace{-2.5ex}
% 	\label{p4_fpga_overall_architecture} 
% \end{figure} 

\begin{figure}
	\centering
    \includegraphics[width=3.5in]{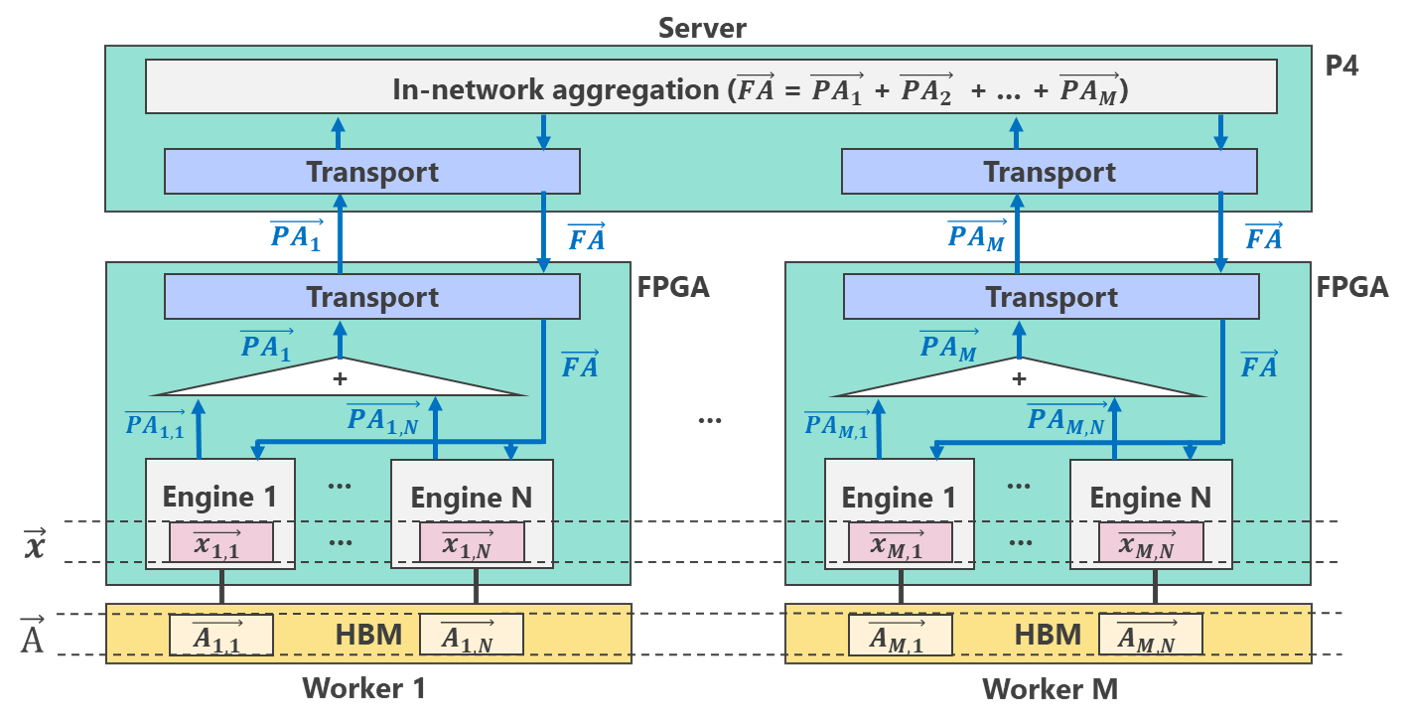}
% 	\subfloat[Detailed design of server]{\includegraphics[width=2.9in]{figure/hw_master.png} \label{server_logic}}	
	%\caption{}
	\vspace{-3ex}	
	\caption{\SystemName{} consists of $M$ workers and one server, and each worker has N engines. \SystemName{} realizes model parallelism between distributed workers, and forward-communication-backward pipeline parallelism within an engine in a worker. As such, \SystemName{} is the first to achieve linear scalability between distributed accelerators.}
	\vspace{-3ex}
	\label{architecture} 
\end{figure}

%\subsection{Design of \SystemName}
\noindent{\bf Overall Architecture. }To achieve the above three goals, we present \SystemName, a P4-switch-enhanced hardware training system that allows efficient scale-out GLM training through model parallelism on distributed FPGAs. In particular, \SystemName{} consists of $M$ FPGA-based workers and one P4 switch enhanced server, 
% \marginpar{R1.O2}
as shown in Figure~\ref{architecture}. All the $M$ workers go through each iteration in a lock step, to efficiently train on the same model $\vv{x}$. Each worker trains on a subset of the model over a subset of the dataset via model parallelism. During each iteration, the $m$-th worker that trains on the model subset $\vv{x_m}$ sends outs a network packet containing its partial aggregation $\vv{PA_m}$ to the P4-switch-based server, which aggregates them and then broadcasts a network packet containing $\vv{FA}$ to all the workers for further backward propagation computing. 

One key idea of \SystemName{} is to divide each mini-batch of samples into multiple smaller micro-batches ({\bf G2}), such that the SGD hardware can explore inter-micro-batch (i.e., intra-mini-batch) parallelism to maximize the overlap between forward/backward propagation and communication, while still preserving the precedence for synchronous SGD (Subsection~\ref{sub_worker}), as shown in \autoref{alg_distributed_sgd}. Another key idea is light-
weight, fault-tolerant, latency-centric in-switch aggregation ({\bf G3})
that directly interplays between a P4 switch and distributed
FPGAs to minimize the latency of AllReduce (Subsection~\ref{sub_p4_collective}).
% \marginpar{R1.O1-4}
At the same time, the communication cost reduces, which can greatly increase the scalability of \SystemName{} ({\bf G1}). %The second key idea of the P4-enhanced server~\cite{transaction_vldb_22, lerner2019cidr} is to provide latency-centric in-network aggregation primitive to serve ultra-low-latency aggregation operations directly to FPGA-based workers (Subsection~\ref{sub_p4_collective}). 
%\SystemName{} has two key innovations that enable to efficiently train GLMs on distributed FPGAs: exploring intra-mini-batch parallelism to overlap forward and backward propagation (), and reconfigurable-switch-enhanced collective operation to minimize network latency.

\begin{algorithm} [t]
	\SetAlFnt{\tiny} \linespread{1.0} \selectfont \caption{\sc Model parallel Training}% on FPGAs
	\label{alg_distributed_sgd}
% 	\SetKwInOut{server}{server}
% 	\SetKwInOut{gWorker}{General worker}	
% 	\SetKwInOut{dWorker}{DLTraining worker}
	\SetKwInOut{Define}{Define}
	\newcommand\mycommfont[1]{\ttfamily\textcolor{blue}{#1}}
    \SetCommentSty{mycommfont}
	
	\begin{footnotesize}
	    
		\Define{
% 		    $S$: number of samples, \\
		    $E$: number of epochs, \\
% 			$\gamma$: learning rate, \\
            $M$: number of workers, \\
			$\vv{A_{i}}[j]$: the $j$-th sample's partition in the $i$-th worker,  \\ 
			$PA_m$: partial activation vector in the $m$-th worker,  \\ %Q_s(\vv{a_i})
 			$b_i$: label value of the $i$-th sample,\\
			$\vv{x_i}$: the partial model vector in the $i$-th worker,\\
			$\gamma$: learning rate.}
			
% 		\server{}
        \textbf{\small{\underline{P4-Switch-based Server:}}}\\
		
		%Load $b$;\\
		Issue $start$ to all workers;\\
		\For{$e = 1$ \KwTo $E$}{ 
		  \For{($i = 0$; $i < S$; $i += MB$)}{ 
		    %\tcc{Stage 1: }
		    Pull $\vv{PA_1}, ..., \vv{PA_M}$ from all the workers; \\
		    $\vv{FA} = \vv{PA_1} + ... + \vv{PA_M}$;\\
		    Push $\vv{FA}$ to all the workers;\\
		  %      $\#$pragma parallel in hardware \\
		  %		\For{($k = 0$; $k < MB$; $k++$)}{
	   %             int32 $\vv{scale}[k] = \gamma \times df(\vv{FA}[k], b_{i+k})$;\\
			 %   }
		  %  Push $\vv{scale}$ to all the workers;\\
		    }
		}
		
% 		\dWorker {m = 1, ..., M}
        \textbf{\small{\underline{FPGA-based worker m (1, ..., M): }}}\\
 		Load the $m$-th partition $\vv{a_m}$ of the training dataset $\vv{a}$;\\
 		Init the $m$-th partition $\vv{x_m}$ of the model $\vv{x}$;\\
\For{$e = 1$ \KwTo $E$}{ 
            
% 			\tcc{Compute $activations$ using local partitions of model and samples}
% 			int32 $activations =\vv{a_{i}} \boldsymbol{\cdot} \vv{x}$;\\
% 			Push $activations$ to server;\\
% 			Pull $scale$ from  server;\\
% 			\tcc{Compute the gradients using the $scale$ and samples}
% 			$\vv{g_{i}} = scale \times \vv{a_{i}}$;\\
% 			Update the local model using the gradients;\\

	\For{($i = 0$; $i < S$; $i += B$)}{
% 	 \tcc{Zero the gradient of this mini-batch}
		$\vv{g_m}$ = 0;	\tcc{Zero the partial gradient}
		\For{($j = 0$; $j < B$; $j += MB$)}{
		    \tcc{Stage 1: forward propagation}%compute $partial\ activations\ (PA)$ using local partitions of model and samples}
		    $\#$pragma parallel in hardware \\
			\For{($k = 0$; $k < MB$; $k++$)}{
				int32 $t = i + j +k$; \\
			    int32 $\vv{PA_m}[k] = \vv{A_{m}}[t] \boldsymbol{\cdot} \vv{x_m}$;\\
			    }
			    \tcc{Stage 2: communication}
			    Push the partial activation vector $\vv{PA_m}$ to the server;\\
			    Pull the full activation vector $\vv{FA}$ from the server;\\
			    \tcc{Stage 3: backward propagation}%compute the gradients using the $scale$ and samples}
			    $\#$pragma parallel in hardware \\
			\For{($k = 0$; $k < MB$; $k++$)}{
			    int32 $t = i + j + k$; \\
			    int32 $\vv{scale}[k] = \gamma \times df(\vv{FA}[k], b_{i+k})$;\\
			    $\vv{g_m} += \vv{scale}[k] \times \vv{A_{m}}[t]$;%\\
			    %$\vv{g_m} = \vv{g_m} + \vv{g_{m,j+k}}$;
			}
		}
        % \tcc{Update the local model using the gradients}
		$\vv{x_m} = \vv{x_m} - \vv{g_m}/B$;
	}			
}	
    \vspace{-1ex}
	\end{footnotesize}
\end{algorithm}

\subsection{Architecture of an FPGA-based Worker}
\label{sub_worker} 
%Model parallelism requires each worker to train a partition of the model on a partition of input dataset, launching a network collective operation after each iteration. Each worker has less computation task per iteration when more workers are involved, making it difficult to amortize negative effect of communication.  

% Inspired by GPipe~\cite{gpipe_NEURIPS2019} that proposes pipeline parallelism across multiple accelerators to accelerate DNN training, we intend to divide a mini-batch of samples into smaller micro-batches when training on the mini-batch.
Algorithm~\ref{alg_distributed_sgd} illustrates the detailed flow of the $m$-th FPGA-based worker under \SystemName{}. In the beginning, it loads its partition of dataset $\vv{a_m}$ (Line 11) and initiates its partition of model $\vv{x_m}$ (Line 12). The algorithm is iterated in $E$ epochs (Line 13). In each epoch, its partition of the input dataset is scanned, one mini-batch of $B$ samples within an iteration (Line 14). Within a mini-batch, we zero the partial gradient $\vv{g_m}$ to 0 (Line 15). Then, it keeps accumulating the gradient $\vv{g_m}$ from a micro-batch ($MB$) of samples, as illustrated in the following three stages.  

%In the following, we describe each of the work 
   
\textbf{Stage 1: Forward Propagation}. The $m$-th worker reads $MB$ partial samples and its partial model $\vv{x_m}$, and computes \emph{partial activations} ($\vv{PA_m}$) of $MB$ elements (Lines 17-21). %Each worker reads training data from memory on FPGA and computes $partial\ activations (PA)$ using its training data and model. The $PA$ is pushed to the server afterward by the network.

\textbf{Stage 2: Communication}. The $m$-th worker sends $\vv{PA_m}$ to the server for aggregation, with the payload of $MB$ elements. Then, it waits for the corresponding \emph{full activation} ($\vv{FA}$) for the future backward propagation (Lines 22-23). %Each AllReduce operation  payload size.
%The server reads label data from HBM or DDR as well as aggregates $PA$ from all workers. Then the server computes the $scale$ using the label value and $full\ activations (FA)$. Last, the server broadcast the $scale$ to all workers.

\textbf{Stage 3: Backward Propagation}. The $m$-th worker computes the partial gradient from the micro-batch of samples and accumulated it to the partial model $\vv{x_m}$ (Lines 25-29). Then, the partial model $\vv{x_m}$ is updated with the average gradient $\vv{g_m}$. %updates the local model using $scale$ received from the server and local data.

    % \vspace{-1ex}
\subsubsection{Parallelism Analysis}  %$\\$
Figure~\ref{mp_microbatch} illustrates the dependency of three stages in the micro-batch pipeline-parallel training within a worker. After the first micro-batch $F_{1,1}$ of the first mini-batch finishes forward propagation computation, it directly enters the communication stage by launching a network collective operation within a micro-batch, without waiting for the other micro-batches to finish. When the first batch $B_{1,1}$ receives the corresponding full activation vector from the server, it can directly enter the third stage. We observe that there is no dependency between micro-batches (e.g., $F_{1,1}$ and $F_{1,2}$) within the same mini-batch, we can overlap communication and forward/backward propagation computation within the mini-batch.% ({\bf Goal G2}). 

\noindent{\bf Discussion. }One potential limitation is that training on a small micro-batch (e.g., 8) of samples at a time would reduce computing parallelism and then under-utilize compute power and network bandwidth on modern CPUs/GPUs, which rely on software implementations. However, \SystemName{} that relies on pure hardware implementation, e.g., FPGA and P4, can more tolerate small micro-batch training, without sacrificing computing throughput.

    % \vspace{-1ex}
\subsubsection{Estimating Time per Iteration under \SystemName} %$\\$
\SystemName{} enables the overlap between forward and backward propagation computation between micro-batches, as shown in Figure~\ref{mp_microbatch}, so we estimate the elapsed time $T_{it}$ per iteration to be forward propagation time $\frac{BW}{B} \times T_{f\_M}$ of a micro-batch plus communication time $T_c$ plus backward propagation time $T_{b\_M}$ of a mini-batch, as shown in Equation~\ref{equ_mp_micro_batch}. Compared with the mini-batch model-parallel training shown in Figure~\ref{mp_overlap}, the proposed micro-batch training is able to maximize the overlap between forward and backward propagation. 
\begin{equation}
\label{equ_mp_micro_batch}
    % T_{it} = \frac{MB}{B} \times T_{f\_M} + T_{c} + T_{b\_M} = \frac{MB}{B} \times T_{f\_M} + T_{b\_M} + \frac{MB}{BW} + T_l
    T_{it} = \frac{MB}{B} \times T_{f\_M} + T_{b\_M} + \frac{MB}{BW} + T_l
\end{equation}

    \vspace{-1ex}
\subsection{P4-Switch-FPGA Collective Operation}
\label{sub_p4_collective}
%Add a figure "" to show the difference between the case with and without p4 switch.
According to the above analysis, communication latency $T_c$ is critical for the overall training time. Furthermore, \SystemName{} needs to perform an \emph{AllReduce} operation on $MB$ elements per iteration, where $MB$ is typically small during our training. Therefore, \SystemName{} requires extremely low latency of an \emph{AllReduce} operation on a small payload. 

Aggregating partial activations in network switches offers two major latency benefits. 
First, as the switch locates inside the network, our approach avoids the additional network hops to and from the aggregation server, essentially reducing activation aggregation to ``sub-RTT (Round Trip Time)" latency.
Second, the switch data plane is designed specifically for fast and predictable packet processing. 
Commercial network switches, including new generation programmable switches~\cite{tofino, tofino2, P4_FPGA_NAI20, Concordia_fast21}, can consistently process packets under a few hundred nanoseconds -- significantly lower even compared to servers with RDMA or kernel-bypassed networking.

However, traditional in-switch aggregation approaches~\cite{sapio2019scaling, lao2021atp}, need end-host servers to prepare packets before in-network aggregation, and thus, unfortunately, incur high communication latency due to unstable software processing overhead and long PCIe latency. Moreover, these approaches introduce a shadow copy mechanism to optimize for high throughput when performing AllReduce on a large message, so these approaches are not friendly to latency due to their late acknowledgement. To this end, we design and implement a light-weight, fault-tolerant, {\bf latency-centric} in-switch aggregation that directly interplays between a P4 switch and distributed FPGAs to minimize the latency of AllReduce needed by each training iteration. % additional network hops, 

\begin{figure}
	\centering
	\includegraphics[width=2.5in]{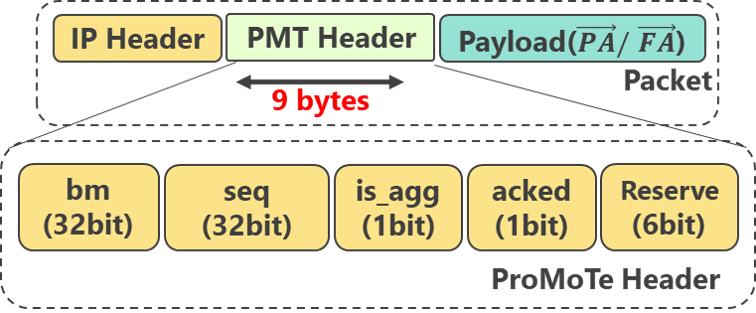} 
	% \caption{}
	\vspace{-0ex}		
	\caption{\SystemName{} Packet Format} 
	\vspace{-3.5ex}
	\label{dlheader} 
\end{figure}

In \SystemName, we handle packet drops by an FPGA-centric retransmission mechanism.
Compared to the approach taken by SwitchML~\cite{sapio2019scaling}\footnote{SwitchML is a widely-used in-switch aggregation method for distributed training. It adopts the shadow copy mechanism to optimize for throughput.}, our solution eliminates the need for shadow copies of the aggregation results on the switch, and significantly reduces switch resource usage.
SwitchML can support half as many outstanding aggregation operations as our approach under the same resource budget.

To address the issue of losing aggregation results due to dropped packets (which contain $\vv{FA}$), our key idea is to use a second round of communication -- initiated and retransmitted by the workers -- to acknowledge the reception of the full activation vector.
Only when the switch receives acknowledgements from all workers for an operation, it can safely clear the aggregation result.

Specifically, as shown in \autoref{alg_switch_unreliable}, the switch only maintains one copy of the aggregated activations, $agg$. 
To separately track the number of received activations and acknowledgement, the switch stores two sets of counters, $agg\_count$ and $ack\_count$.
Since the workers retransmit packets in the event of packet drops, two sets of bitmaps, $agg\_bm$ and $ack\_bm$, are maintained to detect duplicates.
% \marginpar{R2.O3-1}
We also augment the \SystemName{} packet header with additional fields as shown in Figure~\ref{dlheader}: $bm$ is a bitmap with the source worker's index set to one; $seq$ is the aggregation slot index in the P4 switch; $is\_agg$ indicates whether the packet is for aggregation or acknowledgement; $acked$ is a placeholder for the switch to signal that it has received all acknowledgement for the operation.

\begin{algorithm}[!htbp]
    \SetAlFnt{\tiny} \linespread{1.0} \selectfont \caption{\sc Switch aggregation logic with unreliable transmission handling}
    \label{alg_switch_unreliable}
    \SetKwInOut{Initialize}{Initialize}
    \SetKwProg{Fn}{}{:}{end}
    \begin{footnotesize}
        \Initialize{
            $N$: number of aggregation slots \\
            $W$: number of workers \\
            $\vv{agg}[N] := \{\vv{0}\}$ \\
            $agg\_count[N], agg\_bm[N] := \{0\}$ \\ 
            $ack\_count[N], ack\_bm[N] := \{0\}$
        }
        \Fn{\textbf{receive} pkt($is\_agg, seq, bm, \vv{PA}, \vv{FA}$)}
        {
            \If{$is\_agg$}
            {
                \If{$agg\_bm[seq] \& bm = 0$}
                {
                    $agg\_count[seq] \gets agg\_count[seq] + 1$\;
                    $agg\_bm[seq] \gets agg\_bm[seq] | bm$\;
                    $\vv{agg}[seq] \gets \vv{agg}[seq] + pkt.\vv{PA}$\;
                    \If{$agg\_count[seq] = W$}
                    {
                        $ack\_count[seq] \gets 0$\;
                        $ack\_bm[seq] \gets 0$\;
                    }
                }
                \If{$agg\_count[seq] = W$}
                {
                    $pkt.\vv{FA} \gets \vv{agg}[seq]$\;
                    forward $pkt$ to all workers\;
                }
            }
            \Else{
                \If{$ack\_bm[seq] \& bm = 0$}
                {
                    $ack\_count[seq] \gets ack\_count[seq] + 1$\;
                    $ack\_bm[seq] \gets ack\_bm[seq] | bm$\;
                    \If{$ack\_count[seq] = W$}
                    {
                        $agg\_count[seq] \gets 0$\;
                        $agg\_bm[seq] \gets 0$\;
                        $\vv{agg}[seq] \gets \vv{0}$\;
                    }
                }
                \If{$ack\_count[seq] = W$}
                {
                    forward $pkt$ to all workers\;
                }
            }
        }
    \end{footnotesize}
\end{algorithm}

We augment the worker protocol accordingly (\autoref{alg_fpga_unreliable}).
When sending partial activation to the switch, the worker adds its node bitmap to the packet, and indicates that the packet is intended for aggregation (Line 5).
Once the worker receives a full activation, it sends an acknowledgement to the switch (Line 22-23).
However, it only re-enables the slot for aggregation after it receives an acknowledgement confirmation from the switch (Line 26-29).
To handle packet drops, the worker starts a timer after sending each packet (Line 11, 24), and retransmits the packet after the timer expires (Line 31-34).
A timer is canceled once the worker receives the corresponding full activation (Line 20) or acknowledgement confirmation (Line 28).

\begin{algorithm}[h]
    \SetAlFnt{\tiny} \linespread{1.0} \selectfont \caption{\sc Worker-side aggregation logic with unreliable transmission handling}
    \label{alg_fpga_unreliable}
    \SetKwInOut{Initialize}{Initialize}
    \SetKwProg{Fn}{}{:}{end}
    \begin{footnotesize}
        \Initialize{
            $N$: number of aggregation slots \\
            $unused[N] := \{\textsf{true}\}$ \\ 
            $seq := 0$, $bm := \textsf{WORKER\_INDEX}$ 
        }
        \Fn{\textbf{send} pa\_pkt($\vv{PA}$)}
        {
            \If{$unused[seq]$}{
                $unused[seq] \gets \textsf{false}$\;
                $pkt.seq \gets seq; pkt.\vv{PA} \gets \vv{PA}$\; 
                $pkt.bm \gets bm; pkt.is\_agg \gets \textsf{true}$\;
                $seq \gets seq + 1$\;  
                 \If{$seq = N$}{
                    $seq \gets 0$\;
                } 
                forward $pkt$ to switch\;
                start\_timer($pkt$)\;
                return \textsf{true}\;
            } 
            \Else{
                return \textsf{false}\;
            }        
        }
        
        \Fn{\textbf{recvive} pkt($is\_agg, seq, bm, \vv{PA}, \vv{FA}$)}
        {
            \If{$pkt.is\_agg$}{
                cancel\_timer($pkt$)\;
                forward $pkt.\vv{FA}$ to backward propagation\;
                $pkt.is\_agg \gets \textsf{false}; pkt.bm \gets bm$\;
                forward $pkt$ to switch\;
                start\_timer($pkt$)\;
            }
            \Else
            {
                $unused[pkt.seq] \gets \textsf{true}$\;
                cancel\_timer($pkt$)\;
            }
        }
        
        \Fn{\textbf{upon timeout} (pkt)}
        {
            % \If{pkt.isagg}{
            %     $pkt.seq \gets seq; pkt.pa \gets agg[seq]; pkt.isagg \gets 1; pkt.bm \gets bm$\;
            % }
            % \Else
            % {   
            %     $pkt.seq \gets seq; pkt.isagg \gets 1; pkt.bm \gets bm$\;
            % }  
            forward $pkt$ to switch\;
            start\_timer(pkt)\;
        }        
    \end{footnotesize}
\end{algorithm}

% % \input{datastorage.tex}
\vspace{-1.5ex}
\section{Implementation of \SystemName}
%In this section, we present the detailed implementation of \SystemName. In the following, we present the overall hardware architecture, followed by the concrete design an FPGA-based worker and a P4-based server. 
%The implementation of \SystemName{} consists of one server and multiple workers, as shown in Figure~\ref{p4fpga_overall_architecture}. We perform model-parallel training on GLMs. 
%The design of \SystemName{} includes many aspects, such as how to read data from the FPGA’s memory, how the FPGA’s computing logic is, how to transmit data through the network, and so on. We will discuss these designs separately. 

\begin{figure*}[t]
	\centering
% 	\subfloat[Overall architecture of \SystemName, allowing model-parallel training between workers]{\includegraphics[width=5.2in]{figure/hw_design_worker.png} \label{p4fpga_overall_architecture}} 
% 	\subfloat[Detailed design of server]{\includegraphics[width=2.9in]{figure/hw_master.png} \label{server_logic}}	
% 	\hfill
	\subfloat[Detailed design of the $n$-th engine in the $m$-th worker]{\includegraphics[width=3.4in]{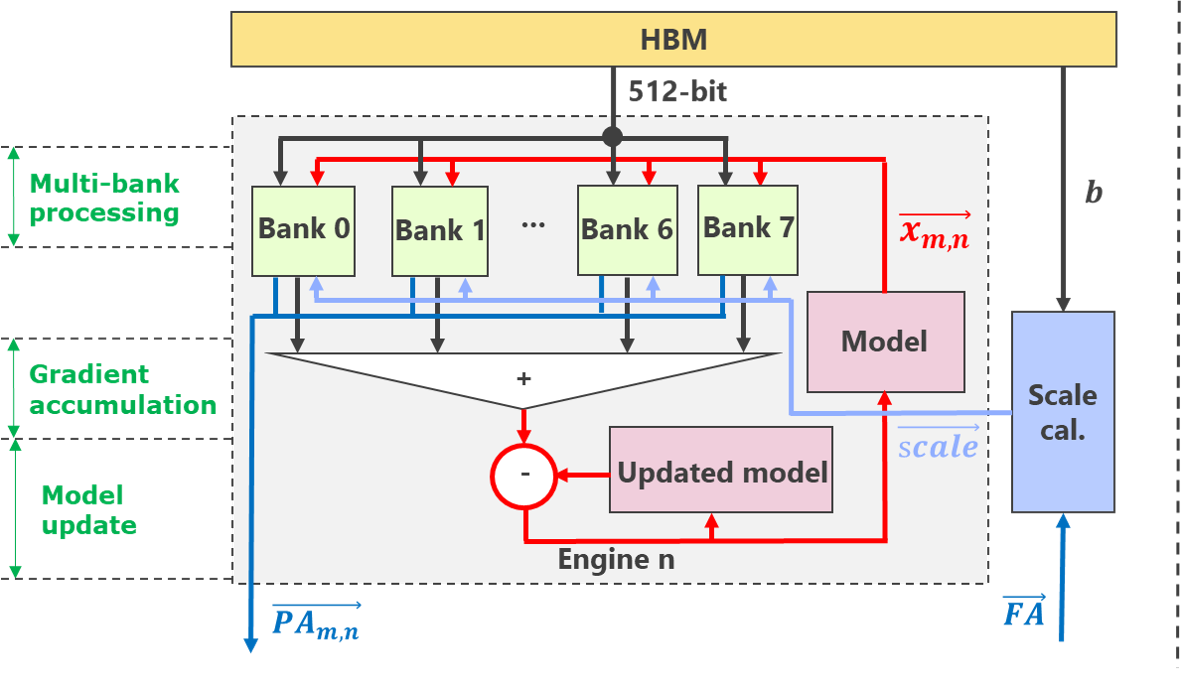} \label{hw_design_f}} 
	\subfloat[Detailed design of the $k$-th bank in the $n$-th engine in the $m$-th worker, allowing forward-communication-backward pipeline parallelism]{\includegraphics[width=2.6in]{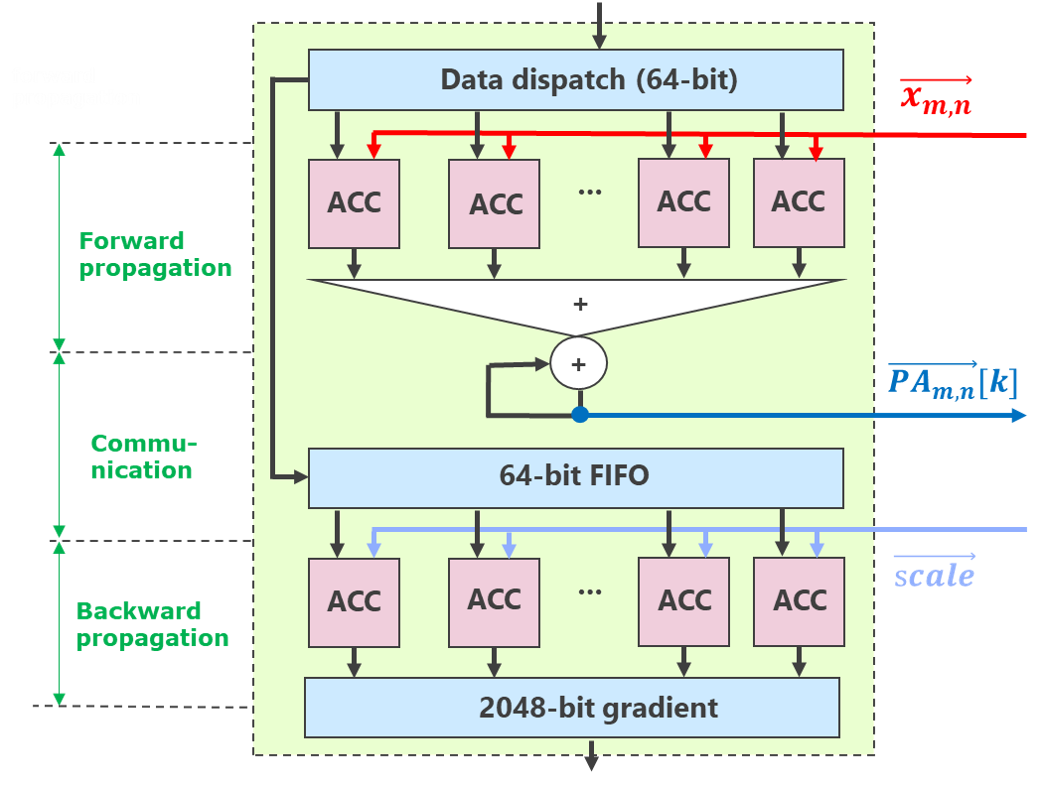} \label{hw_design_b}}	
	%\caption{}
	\vspace{-1ex}	
	\caption{Each worker of \SystemName{} has N engines. Each engine has 8 banks to populate a micro-batch of 8 samples.}
	\vspace{-3ex}
	\label{implement} 
\end{figure*}

%\subsection{Main Architecture of \SystemName}
We implement \SystemName{} with one P4-switch-based server and $M$ FPGA-based workers. In the following, we present the implementation details of an FPGA-based worker and P4-switch-based server, as shown in Figure~\ref{implement}. %We perform model-parallel training on GLMs. 
%\subsection{Overall Hardware Architecture of \SystemName{}}

% The architecture of \SystemName{} has many levels. A complete \SystemName{} is composed of multiple workers and a server. Each worker is consists of multiple engines, and in each engine, the calculation pipeline is divided into multiple banks. We introduce the whole \SystemName{} from high level to details.

% Figure~\ref{p4fpga_overall_architecture} illustrates the overall architecture of \SystemName{}. \SystemName{} consists of M workers and one server. Each worker is located on an FPGA board. This FPGA board is connected to the CPU through PCIe. The CPU does not participate in the calculation process of distributed training, but only controls the \SystemName{}, displays the results, and sending the data to the memory of FPGA before the training starts.

% There are N engines in each worker. After the engine calculates $PA$, it sends $PA$ to the server via network module, and waits for the server to return the $scale$ or $FA$, and then continues the SGD calculation.

% The server is implemented on FPGA or P4 switch. there are no engines in the server, instead of an $Agg\ a$ module, which adds up the $PA$ generated by all the engines in the workers, and sends the $scale$ or $FA$ to all the workers.

    \vspace{-1ex}
\subsection{Hardware Design of a Worker}%n FPGA-based
\label{worker_hw_design}
The goal of the FPGA-based worker is three-fold. First, it maximizes the processing parallelism ({\bf E1}), which needs not only high processing ability but also high memory bandwidth required by GLM training. Second, it enables micro-batch pipeline-parallel training to overlap forward and backward propagation computation to increase computing pipeline utilization ({\bf E2}). Third, it needs to overlap communication and forward/backward propagation computation to amortize the negative effect of inter-FPGA communication ({\bf E3}). 

We implement each worker with verilog language on an HBM-equipped FPGA board Xilinx Alveo U280~\cite{u280}.
% , which also features a large number of logics, DSPs, and on-chip MEMs. 
In order to achieve {\bf E1}, we adopt a multi-engine design to partition a subset of the model assigned to this worker uniformly to $N$ engines, where N is parameterized at compile time. As such, all the engines train in a lock step and each engine only needs to train a small portion of the subset. For example, $\vv{x_{1,2}}$ corresponds to the model portion associated with the second engine of the first worker.\footnote{We also vertically partition the dataset $\vv{A}$ in the same way. } Each model portion, accommodating up to 256K weights,\footnote{This number is parameterizable at compile time, under the constraint of FPGA resource limitation. However, we can easily generalize \SystemName{} to support a large model that is stored in external memory, e.g., HBM, without affecting performance. } is implemented with on-chip memory, so each worker supports 2M weights.  Therefore, each worker consists of an HBM subsystem (subsection~\ref{subsub_hbm}), $N$ engines (subsection~\ref{subsub_engine},~\ref{subsub_bank}),
% \footnote{We put the hardware details of each engine in Appendix due to page limitation. }
% and a reliable transport (Subsection~\ref{subsub_transport}), 
as shown in Figure~\ref{architecture}.

% \hhj{The FPGA-based worker is connected to the CPU via a PCIe interface. The CPU does not actively participate in the distributed training process, but it controls the initiation of P4SGD via the PCIe MMIO interface and downloads the trained model from the FPGA using the PCIe DMA interface.}

% 	\vspace{-1ex}
\subsubsection{Hardware Design of Memory Subsystem}\label{subsub_hbm} %$\\$  
\begin{figure}[t]
	\centering
	\includegraphics[width=3.3in]{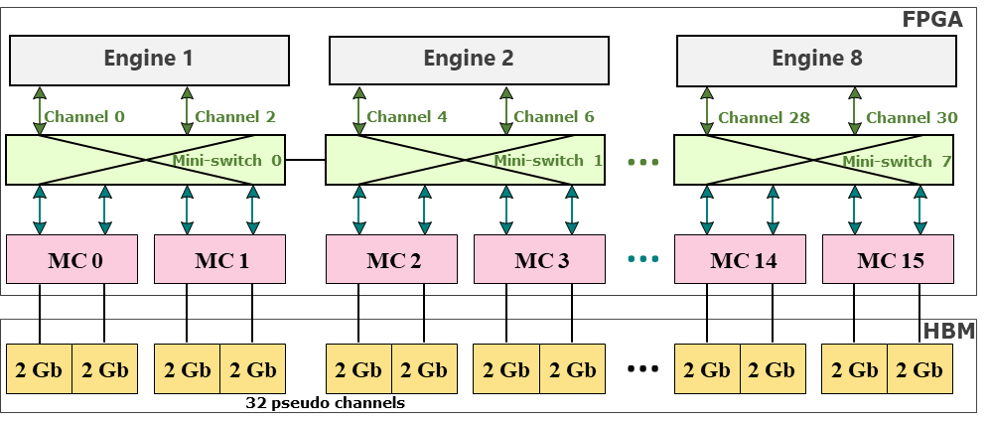} 
	% \caption{}
	\vspace{-1ex}		
	\caption{Relationship between 16 HBM channels and 8 engines, each engine occupies 4 HBM pseudo channels.} 
	\vspace{-2.5ex}
	\label{hbm_to_engine} 
\end{figure}

The memory subsystem employs the HBM subsystem that has 32 memory channels to support 32 independent 256-bit memory accesses. Due to FPGA resource limitations, an FPGA board can only afford to instantiate up to $N$=8 engines, each engine can occupy four consecutive HBM channels to provide sufficient memory bandwidth and capacity (8Gb) for its subset of the dataset. Figure~\ref{hbm_to_engine} illustrates the relationship between 32 HBM channels and 8 engines. Since each engine is designed to process 512-bit data stream per cycle, each engine combines two 256-bit AXI interfaces to access the data in the HBM. %each engine can read 512 bits of data for calculation in each cycle.

%$MB$ micro-batch...
%one bank for one sample...
%64 features concurrently...
%model is in the on-chip memory...

\subsubsection{Hardware Design of an Engine} \label{subsub_engine} %$\\    $ 
%The key goal of an engine is to train . 
Built on the state-of-the-art GLM accelerator MLWeaving~\cite{mlweaving2019vldb} that supports efficient low-precision training, each engine extends to enable GLM training on distributed FPGAs in a lock step.
To achieve {\bf E2} in section~\ref{worker_hw_design},
% Since each worker needs to support micro-batch model-parallel training on GLM, 
we adopt a multi-bank design to allow each engine to train on a micro-batch of $MB$=8 samples concurrently, one bank to populate a sample. As such, $MB$ samples flow into the computing pipeline concurrently. 

The hardware design of each engine consists of three stages: ``multi-bank dividing”, ``gradient accumulation”, and ``model update”.
% \marginpar{R2.O3-1} 
In the “multi-bank dividing” stage, the 512-bit input data stream flows into 8 banks, each of which consumes 64-bit data from the same sample, one bit from one feature of a sample. We leave the detailed design of each in Subsection~\ref{subsub_bank}. The output of each bank is a 2048-bit gradient (i.e., 64 32-bit gradient elements from 64 features). 
In the “gradient accumulation” stage, we instantiate 64 8-element-wise adder trees to aggregate gradients from all  8 banks. Unlike MLWeaving, we use DSP instead of LUT to construct the adder tree. Each level of the adder tree can add three numbers, which saves resources and reduces calculating time.
In the “model update” stage, the aggregated gradient from the above $MB$=8 samples is accumulated into the ``updated model". When accumulating the gradient from the last micro-batch of each mini-batch into the ``updated model", we also update the architectural model $(\vv{x})$ with the same value written into the ``updated model".

% The hardware design of each engine consists of three stages: ``multi-bank dividing”, ``gradient accumulation”, and ``model update”. 
% In the following, we will present the detailed hardware design of each stage.

% In the “multi-bank dividing” stage, the 512-bit input data stream flows into into 8 banks, each of which consumes 64-bit data from the same sample, one bit from one feature of a sample. We leave the detailed design of each in Subsection~\ref{subsub_bank}. The output of each bank is 2048-bit gradient (i.e., 64 32-bit gradient elements from 64 features).  

% In the “gradient accumulation” stage, we instantiate 64 8-element-wise adder trees to aggregate gradients from all the 8 banks. Each adder tree consumes one 32-bit gradient element from each of eight banks, and produces one aggregated gradient element. As such, it generates 64 elements of the aggregated gradient within a cycle.

% In the “model update” stage, the aggregated gradient from the above $MB$=8 samples is accumulated into the ``updated model". When accumulating the gradient from the last micro-batch of each mini-batch into the ``updated model", we also update the architectural model $(\vv{x})$ with the same value written into the ``updated model". 

% 	\vspace{-0.5ex}
\subsubsection{Hardware Design of a Bank} \label{subsub_bank} %$\\$
%According to Algorithm 
Figure~\ref{hw_design_b} shows the details of each bank. 
According to the distributed SGD algorithm, each bank consists of three stages.
To achieve {\bf E3} in section~\ref{worker_hw_design}, we send “partial activations" of $MB$ samples each time to support micro-batch model-parallel training on GLM in the "communication" stage.

The hardware design of each bank consists of three stages: ``forward propagation”, ``communication”, and ``backward propagation”. In the ``forward propagation" stage, we instantiate 64 bit-serial multipliers~\cite{eckert2018neural,umuroglu2018bismo} to consume 64-bit data stream, a bit from each feature. 
% \marginpar{R2.O3-1}
 At the same time, the 64-bit data stream is also fed into the 64-bit ``FIFO". Each bit-serial multiplier outputs a 32-bit calculation result, which is directly fed into the full-pipelined adder tree. The output of the adder tree feeds to an accumulator, which aggregates the corresponding ``partial activation".  
 In the ``communication" stage, together with ``partial activations" from the other banks in the same engine, the $n$-th engine prepares $\vv{PA_{m,n}}$ for the current micro-batch of $MB$ samples. The $m$-th worker then aggregates ``partial activations" from all its $N$ engines to produce $\vv{PA_{m}}$, which is sent to the P4-switch-based server for in-network aggregation in Figure~\ref{architecture}.   
 In the ``backward propagation" stage, we also instantiate 64 bit-serial multipliers, each of which reads a 1-bit feature from the 64-bit ``FIFO" and multiplies it with the corresponding 32-bit element $\vv{scale}[k]$ per cycle. Therefore, 64 bit-serial multipliers is able to produce 2048-bit gradient computation results per cycle for further processing in the corresponding engine (Subsection~\ref{subsub_engine}).

\begin{figure}
	\centering
	\includegraphics[width=3in]{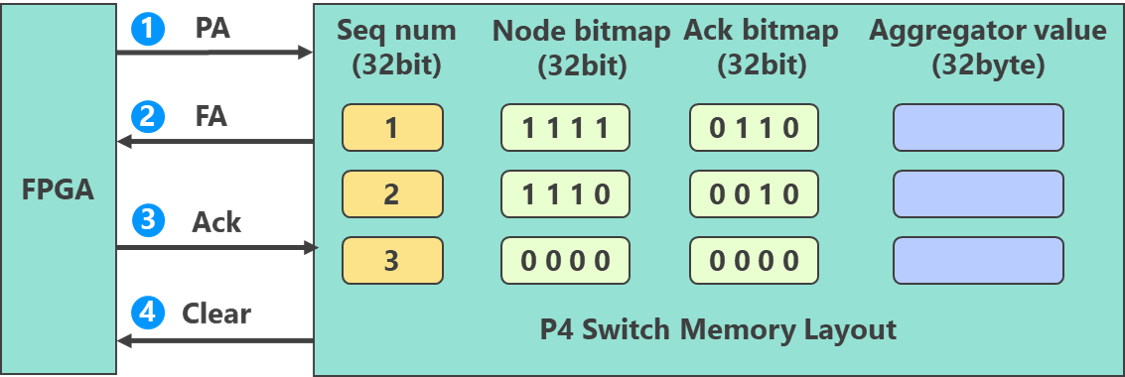} 
	% \caption{}
	\vspace{-1ex}		
	\caption{Interaction between P4 switch and FPGA} 
	\vspace{-1.5ex}
	\label{fig_nteraction} 
\end{figure}

% \begin{figure}
% 	\centering
% 	\includegraphics[width=3in]{figure/dl_header.png} 
% 	% \caption{}
% 	\vspace{-0ex}		
% 	\caption{\SystemName{} Packet Format} 
% 	\vspace{-3.5ex}
% 	\label{dlheader} 
% \end{figure}

% \begin{figure}
% 	\centering
% 	\includegraphics[width=2.7in]{figure/hw_transport.png} 
% 	% \caption{}
% 	\vspace{0ex}		
% 	\caption{Detailed design of the transport on the FPGA} 
% 	\vspace{0ex}
% 	\label{hw_transport} 
% \end{figure}

% 	\vspace{-1.5ex}
% \subsection{Implementation of Reliable Transport}
\vspace{-1.5ex}
\subsection{Implementation of P4-switch-based Server} %$\\$\label{subsub_transport}

Figure~\ref{fig_nteraction} briefly describes the hardware implementation of reliable transport between P4 switch and an FPGA. %In the following, we present the detailed implementation of each side. 

Our switch data plane is implemented in the P4 language. 
Each of the $agg$, $agg\_count$, $agg\_bm$, $ack\_count$, and $ack\_bm$ in \autoref{alg_switch_unreliable} is mapped directly to a Tofino register array, as shown in Figure~\ref{fig_nteraction}.
In our current configuration, the size of the register arrays is set to 64K (16 bits indices), permitting a maximum of 64K outstanding aggregation operations.
The register arrays are distributed over 4 stages of a switch pipeline -- out of the 12 total stages available.
Resource consumption in any of the 4 stages is capped at $70.83\%$ of the available SRAM.
This leaves ample resources for the other bread-and-butter switching functionalities.
We leverage the Tofino packet replication engine to implement multicasting to workers.

\section{Evaluation}

\subsection{Experimental Setting}

\noindent{\bf Workloads. } We conduct our experiments on five datasets with the various number of features, as shown in Table~\ref{t_dataset}. All datasets used are publicly accessible. 
\begin{table} [t]
	\centering
    \vspace{-1ex}
	%\begin{spacing}{0.3}
	\begin{scriptsize}
	\caption{Evaluated datasets }
	\label{t_dataset}
		\vspace{-1.5ex}
		\begin{tabular}{|c||c|c|c|}
			\hline
			\textbf{Dataset} & \textbf{Samples} & \textbf{Features}  & \textbf{Classes} \\%testing samples\\
			\hline			\hline
			gisette\cite{dataset_gisette} & 6,000 &  5,000  & 2\\
			\hline
			real\_sim\footnotemark[7] & 72,309 &  20,958  & 2\\
			\hline
			rcv1\footnotemark[7] & 20,242 & 47,236 &2\\
			\hline
			amazon\_fashion\footnotemark[8] & 200,000 & 332,710 & 5\\
			\hline
			avazu\footnotemark[7] & 40,428,967 & 1,000,000 & 2\\			
			\hline
		\end{tabular}
	\end{scriptsize}
	\vspace{-2ex}
	%\end{spacing}\begin{tabular}{@{}c@{}}ImageNet \\ (Transfer Learning)\end{tabular} $\#$  $\#$ $\#$ $\#$ 
\end{table}

\addtocounter{footnote}{0} %3=n
 \stepcounter{footnote}\footnotetext{https://www.csie.ntu.edu.tw/$\sim$cjlin/libsvmtools/datasets/}
 \stepcounter{footnote}\footnotetext{This is part of the indian news dataset, and the dataset can be downloaded from https://jmcauley.ucsd.edu/data/amazon/}

\noindent{\bf Experimental Platform.} We run our experiments on a cluster with two network switches and eight machines. One network switch is Mellanox SN2700 with 32 x QSFP28 ports, and the other one is the Wedge100BF-32X P4 reconfigurable switch which provides 32 x QSFP28 ports and a 20 MB packet buffer. Each machine configured with a 12-core/24-thread Intel Xeon(R) Silver 4214 CPU (2.2GHz), a Xilinx Alveo U280 FPGA~\cite{u280}, a 100Gb/s (RDMA-enabled) Mellanox MT27800 NIC and a 
% \marginpar{R2.O4-1}
NVIDIA A100 GPU (40 GB HBM2 memory, 6912 CUDA cores).
% \footnote{We use 4 NVIDIA A100 (40 GB HBM2 memory, 6912 CUDA cores) and 4 NVIDIA V100 (32 GB HBM2 memory, 5120 CUDA cores). Each machine has exactly one of these GPUs.}
We use this platform for all the experiments. 
%We use a cluster with 8 machines and 2 network switches in our experiments, each machine configured with 1 CPU, 1 GPU, 1 FPGA and 1 network card. One of the network switches is Mellanox SN2700 with 32 x QSFP28 ports, the other one is the P4 switch. We use this cluster for all the experiments. We don't have eight same GPUs, so we use 2 NVIDIA Quadro RTX 8000 (48 GB GDDR6 memory, 4608 CUDA cores), 2 NVIDIA A100 (40 GB HBM2 memory, 6912 CUDA cores) and 4 NVIDIA V100 (32 GB HBM2 memory, 5120 CUDA cores). Each machine has exactly one of these GPUs.

% {\bf P4 Switch-based Server Implementations.} We run the P4 switch-based server in Wedge100BF-32X which . 

\noindent{\bf \SystemName{} Implementations. } 
We implement \SystemName{} with a P4-switch-based server and FPGA-based workers. 
%  We adopt open-source TCP stack~\cite{20.500.11850/362532} as our communication transport between the server and workers.  
Table~\ref{t_resource_consumption} shows the resource consumption of \SystemName{} with 8 engines on Xilinx Alveo U280. 
The total resource utilization is about 50$\%$. \SystemName{} allows instantiating a flexible number of engines from 1 to 8. 
Furthermore, we also implement the data-parallel training system that also leverages a P4 switch to do in-network aggregation between distributed FPGAs. The data-parallel system aggregates gradients of length $D$, rather than $B$, after forward and backward propagation within an iteration. The adopted precision is 4 bits because 1) the execution time will decrease linearly as the precision decreases, and 2) MLWeaving~\cite{mlweaving2019vldb} demonstrates that low-precision (above 3 bits) training takes a similar number of epochs to converge as that of full-precision CPU approaches. 
So we choose 4-bits precision to decrease the training time without affecting convergence rate.

%\hhj{In addition, to compare the performance of data-parallel and model-parallel, we write the code for data-parallel training based on \SystemName{} and simulated it, and the comparison results are in section~\ref{da_mo_compare}}

% We implement \SystemName{} with two implementations: "FPGA+P4" and "FPGA-only". 
% "FPGA+P4" represents the case with a P4-switch-based server and FPGA-based workers. 
% "FPGA-only" represents the case with an FPGA-based server and FPGA-based workers. We adopt open-source TCP stack~\cite{20.500.11850/362532} as our communication transport between the server and workers. We introduce "FPGA-only" to examine the effect of in-network aggregation (Subsection~\ref{sub_p4_collective}). 
% Table~\ref{t_resource_consumption} shows the resource consumption of \SystemName{} with 8 engines on Xilinx Alveo U280. 
% The total resource utilization is about 50$\%$. \SystemName{} allows to instantiate flexible number of engines from 1 to 8. 

\begin{table} [t]
	\centering
	%\begin{spacing}{0.3}
		\begin{scriptsize}
	\vspace{-0.5ex}
	\caption{Resource consumption of a worker with 8 engines}% under \SystemName{}
	\label{t_resource_consumption}
	\vspace{-1ex}
	\begin{tabular}{|c||c|c|c|c|c|}
		\hline
		{\bf Hardware modules} &  {\bf LUTs} &  {\bf REGs} &  {\bf RAMs} &  {\bf DSPs}&  {\bf Freq.}\\
		\hline
		\hline
% 		{\bf An engine} & 23.5k & 113k & 19Mb & 512 & 250MHz\\ 
% 		\hline		
		{\bf PCI-Express} & 63K & 98K & 4.3Mb & 0 & 250MHz\\ 
		\hline
% 		{\bf TCP stack} & 46k & 50k & 3.5Mb & 0 & 250MHz\\ 
% 		\hline
		{\bf Network transport} & 10K & 27K & 3.5Mb & 0 & 250MHz\\ 
		\hline		
		{\bf HBM subsystem} & 7K & 42K & 3.26Mb &  0 &450MHz \\ 
		\hline		
% 		\hline
		{\bf 8 engines } & 188K  & 904K & 152Mb & 4096 & 250MHz \\ 
		\hline
% 		{Total resources used } & 304k  & 1094k & 163Mb & 4096 &  \\ 
% 		\hline		
		{\bf Total utilization} & \makecell{{304K} \\ {(23$\%$)}} & \makecell{{1.1M} \\ {(42$\%$)}}  & \makecell{{165Mb} \\ {(47.5$\%$)}}   &  \makecell{{4096} \\ {(45$\%$)}}  & \\

		\hline
	\end{tabular}
	\vspace{-2ex}

		\end{scriptsize}
	%\end{spacing}
\end{table}

%\hhj{\noindent{\bf FPGA Baseline. } We adopt MLWeaving~\cite{mlweaving2019vldb} as our FPGA baseline. It is the state-of-the-art GLMs accelerator on FPGA. We implement MLWeaving on Xilinx Alveo U280.  }

\noindent{\bf GPU Baseline. } The GPU baseline, labeled ``GPUSync", is implemented on the GPUs in the cluster adopting a synchronous distributed linear model SGD. 
% According to Algorithm~\ref{alg_sgd_flow_bank}, model reading (Line 6) and model update (Line 11) has a Read After Write (RAW) dependency, due to the inherently sequential nature of synchronous SGD. Therefore, parallelism exists within a mini-batch of samples. 
``GPUSync" leverages the state-of-the-art cuBLAS library~\cite{cublas} to efficiently implement forward and backward propagation. In particular, ``GPUSync" uses the function \emph{cublasSgemm}, which does auto parallelization within a single GPU. From the NVIDIA Nsight Systems, we observe that \emph{cublasSgemm} uses at least 512 thread blocks, 128 threads per thread block, to compute on a mini-batch of samples in our experiment, indicating that ``GPUSync" has already fully utilized GPU computing power within a CUDA call. Moreover, we leverage a few optimization methods on GPU to accelerate the training: CUDA Graphs~\cite{cudagraph} to reduce kernel invocation overhead, and RDMA+GPUDirect-enabled NCCL~\cite{gpudirect, nccl} to reduce inter-node communication overhead. ``GPUSync" adopts model-parallel training, which is obviously faster than data-parallel training (Subsection~\ref{da_mo_compare}). 
%\hhj{since the data-parallel training is slower than it }
However, ``GPUSync" cannot efficiently scale out due to its CUDA kernel invocation overhead, in particular, each training iteration needs to launch three CUDA kernels: two \emph{cublasSgemm} for forward/backward passes and one AllReduce for communication. When employing more GPUs, the GPU compute cycles per kernel are reduced and kernel invocation overhead can dominates the overall time. % due to communication overhead. 

\noindent{\bf Two CPU Baselines. } We implement the synchronous SGD algorithm on distributed CPUs, labeled ``CPUSync". We employ the following optimization methods on distributed CPUs: multi-core (12 cores), AVX2 instruction (512-bit) and RDMA-based openMPI (version 3.4.1) library. ``CPUSync" adopts model-parallel training, which is obviously faster than data-parallel training in our experiment. 
% CPUSync can efficiently scale out due to linear computation time reduction per iteration, so TCP performance is not the bottleneck. 
The other CPU baseline is ``SwitchML" that adopts the same computation method as ``CPUSync", but uses the communication method from SwitchML~\cite{sapio2019scaling}, rather than RDMA-based OpenMPI. 
%\hhj{Moreover, We measure the aggregation delay of SwitchML, labeled ``SwitchML", to compare with our approach.}
% Moreover, we implement a CPU-based in-switch aggregation, labeled ``CPU+P4", to measure the aggregation latency between distributed CPUs using P4 switch~\cite{sapio2019scaling}. 

\noindent{\bf Comparison Methodology. } %Our evaluations mainly explain the advantages of \SystemName{} from three aspects. First, we compare the network communication delay between \SystemName{} and NCCL, \SystemName{} has a better performance in network data transmission. Second, we examine the hardware efficiency of \SystemName{} in terms of measuring its performance with respect to datasets, mini-batch size, engine size and worker size. When we use $local server$,\SystemName{} can achieve linear speedup as the number of engines increases. When we use $remote server$, \SystemName{} also has a good performance. Third, \SystemName{} converges faster than all baseline systems. 
Our evaluations mainly validate three hypotheses. 
First, P4-switch-FPGA in-network aggregation can significantly reduce the aggregation latency, benefiting model-parallel training on GLMs. %First, Our FPGA-P4 based AllReduce operation is able to achieve extremely low latency,  that heavily relies on low latency between distributed models. 
Second, \SystemName{} is able to achieve an almost linear scale-out on distributed FPGAs, with the help of the P4 switch.
Third, \SystemName{} converges faster and consumes lower than its counterparts on distributed GPUs/CPUs.

% \vspace{-1.5ex}
\subsection{Comparison of Aggregation Latency }
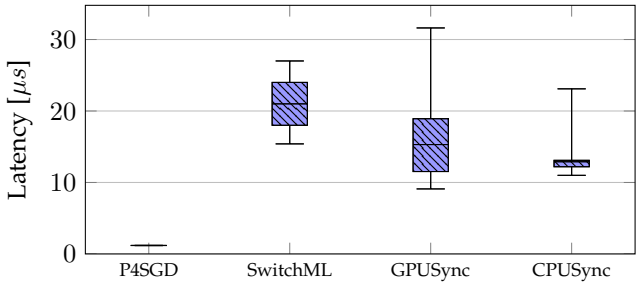
\begin{figure}[h]
\centering
  \begin{tikzpicture}
  
  \begin{axis}[
  	boxplot/draw direction=y,
    width=1\columnwidth,
	height=0.55\columnwidth,
    xtick={0, 2, 4, 6},
    xticklabels={\scriptsize{P4SGD}, \scriptsize{SwitchML}, \scriptsize{GPUSync},\scriptsize{CPUSync}},%{\tiny{R8K\_RD\_FPGA}, \tiny{A100\_RD\_FPGA},\tiny{CPU\_RD\_FPGA},\tiny{R8K\_RD\_CPU},\tiny{A100\_RD\_CPU}},
    ylabel={Latency [$\mu s$]},
    ymin=0.0,
    ymajorgrids=true,
    %symbolic x coords={4, 8, 16, 32},
    %box extend=0.5,
    legend style={at={(0.5,1.05)},
          anchor=south, font=\footnotesize},
    legend columns=3,
    legend cell align=left,
  ]
  \pgfplotstableread{data/latency.dat}\datatable
%   \pgfplotstablegetrowsof{\datatable}
%   \pgfmathtruncatemacro\TotalRows{\pgfplotsretval-1}
  \pgfplotsinvokeforeach{0}
  {
  	\addplot[
    	boxplot prepared from table={
      		table=\datatable, row=#1,
      		lower whisker=1p, upper whisker=99p, lower quartile=25p, upper quartile=75p, median=median, draw position=pos
        },
    	boxplot prepared={ box extend=0.5, },
    	fill=red!40,
    	postaction={pattern=crosshatch},
    	area legend
    ] coordinates {};
  	\addplot[
    	boxplot prepared from table={
      		table=\datatable, row=#1,
      		lower whisker=1p, upper whisker=99p, lower quartile=25p, upper quartile=75p, median=median, draw position=pos
        },
    	boxplot prepared={ box extend=0.5, },
    	pattern=crosshatch,
    	forget plot
    ] coordinates {};

  }
    \pgfplotsinvokeforeach{1}
  {
  	\addplot[
    	boxplot prepared from table={
      		table=\datatable, row=#1,
      		lower whisker=1p, upper whisker=99p, lower quartile=25p, upper quartile=75p, median=median, draw position=pos
        },
    	boxplot prepared={ box extend=0.5, },
    	fill=blue!40,
    	postaction={pattern=north west lines},
    	area legend
    ] coordinates {};
  	\addplot[
    	boxplot prepared from table={
      		table=\datatable, row=#1,
      		lower whisker=1p, upper whisker=99p, lower quartile=25p, upper quartile=75p, median=median, draw position=pos
        },
    	boxplot prepared={ box extend=0.5, },
    	pattern=north west lines,
    	forget plot
    ] coordinates {};

  }
  
    \pgfplotsinvokeforeach{2}
  {
  	\addplot[
    	boxplot prepared from table={
      		table=\datatable, row=#1,
      		lower whisker=1p, upper whisker=99p, lower quartile=25p, upper quartile=75p, median=median, draw position=pos
        },
    	boxplot prepared={ box extend=0.5, },
    	fill=blue!40,
    	postaction={pattern=north west lines},
    	area legend
    ] coordinates {};
  	\addplot[
    	boxplot prepared from table={
      		table=\datatable, row=#1,
      		lower whisker=1p, upper whisker=99p, lower quartile=25p, upper quartile=75p, median=median, draw position=pos
        },
    	boxplot prepared={ box extend=0.5, },
    	pattern=north west lines,
    	forget plot
    ] coordinates {};

  }  

    \pgfplotsinvokeforeach{3}
  {
  	\addplot[
    	boxplot prepared from table={
      		table=\datatable, row=#1,
      		lower whisker=1p, upper whisker=99p, lower quartile=25p, upper quartile=75p, median=median, draw position=pos
        },
    	boxplot prepared={ box extend=0.5, },
    	fill=blue!40,
    	postaction={pattern=north west lines},
    	area legend
    ] coordinates {};
  	\addplot[
    	boxplot prepared from table={
      		table=\datatable, row=#1,
      		lower whisker=1p, upper whisker=99p, lower quartile=25p, upper quartile=75p, median=median, draw position=pos
        },
    	boxplot prepared={ box extend=0.5, },
    	pattern=north west lines,
    	forget plot
    ] coordinates {};

  }

%   \pgfplotsinvokeforeach{4}
%   {
%     	\addplot[
%     	boxplot prepared from table={
%       		table=\datatable, row=#1,
%       		lower whisker=1p, upper whisker=99p, lower quartile=25p, upper quartile=75p, median=median, draw position=pos
%         },
%     	boxplot prepared={ box extend=0.5, }, black, pattern=dots, area legend
%     ] coordinates {};

%   }

  \end{axis}
  \end{tikzpicture}
  \vspace{-1ex}
  \caption{Aggregation latency comparison. Whiskers show the 1st and 99th percentile. }%A:R8K RD FPGA; B:A100 RD FPGA; C:CPU RD FPGA; D:R8K R D CPU; A100 RD CPU
  \vspace{-2ex}
  \label{fig_control_path_latency}
\end{figure}

We measure the latency of P4-switch-FPGA in-network aggregation that serves distributed GLM training on FPGAs (Subsection~\ref{sub_p4_collective}). Figure~\ref{fig_control_path_latency} illustrates the latency comparison of \texttt{AllReduce} that performs an array $\vv{PA}$ of 8 32-bit elements in each of 8 workers.   %result with different devices.
We have three observations. 
%Moreover, We measure the aggregation delay of SwitchML, labeled ``SwitchML", to compare with our approach.}

First, \SystemName{} is able to reach average latency of 1.2$\mu$s, which is an order of magnitude smaller than that of ``CPUSync" and ``GPUSync", because in-switch aggregation reduces 
% the additional network hops for the AllReduce operation. 
the additional network hops, and our hardware implementation reduces software launching and synchronization overhead. %due to its hardware implementation and

Second, the latency fluctuation of \SystemName{} is significantly smaller than that of ``CPUSync" and ``GPUSync", demonstrating one of the advantages of \SystemName{} in terms of offering deterministic latency. 
Third, SwitchML~\cite{sapio2019scaling} introduces longer latency even than ``CPUSync" and ``GPUSync", because 1) SwitchML leverages the shadow copy mechanism to delay the acknowledgment of received aggregation packets for higher throughput, and 2) SwitchML uses data packets with a minimum size of 256B, while other methods adopt 64B network packets.
% Third, \SystemName{} achieves much less latency than``CPU+P4" that leverages SwitchML~\cite{sapio2019scaling} to enable in-switch aggregation to accelerate aggregation between distributed CPUs, indicating that FPGA-based transport is vital to extremely low latency.  

% First, ``FPGA+P4" is able to reach average latency of 1.2$\mu$s, which is significantly smaller than that (4.5$\mu$s) of ``FPGA-only", which implements aggregation on an FPGA that connected to an normal switch, because in-switch aggregation reduces the additional network hops for the AllReduce operation. Moreover, it is an order of magnitude smaller than that of ``GPUSync" and ``GPUSync", due to its hardware implementation.
% Second, the latency fluctuation of ``FPGA+P4" is significantly smaller than that of ``CPUSync" and ``GPUSync", demonstrating one of the advantages of \SystemName{} in terms of offering deterministic latency. 
% Third, ``FPGA+P4" achieves much less latency than``CPU+P4" that leverages SwitchML~\cite{sapio2019scaling} to enable in-switch aggregation to accelerate aggregation between distributed CPUs, indicating that FPGA-based transport is vital to extremely low latency.  
% -----------------------------------------------------------------------------
% -----------------------------------------------------------------------------

\vspace{-1ex}
\subsection{Comparison of Data Parallelism and Model Parallelism}\label{da_mo_compare}
\begin{figure}[t]
	\centering
	\subfloat[$rcv1$ (47K)]{\includegraphics[width=1.8in]{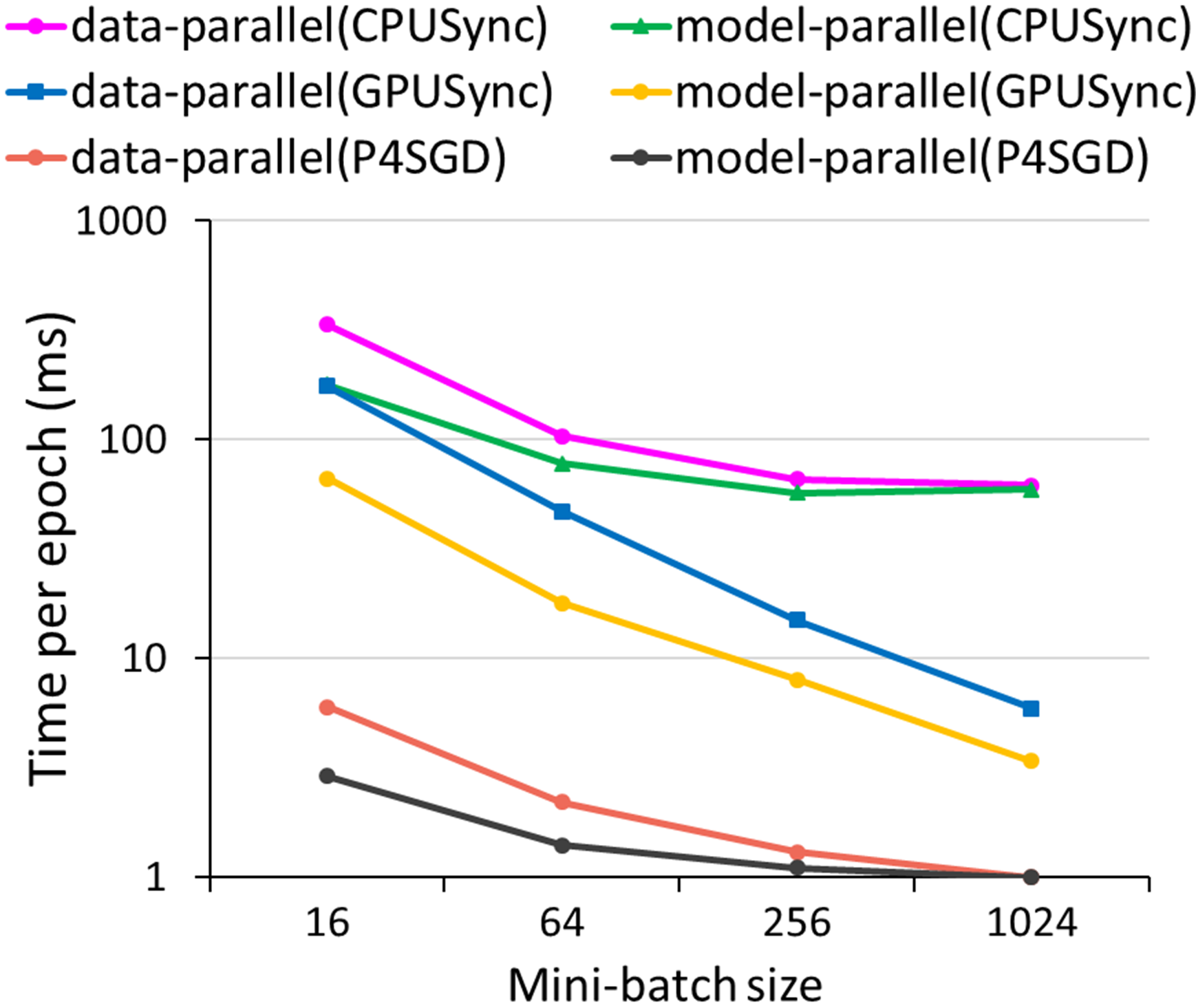} \label{real_data_model_comp}} 
	\subfloat[$amazon\_fasion$ (333K)]{\includegraphics[width=1.8in]{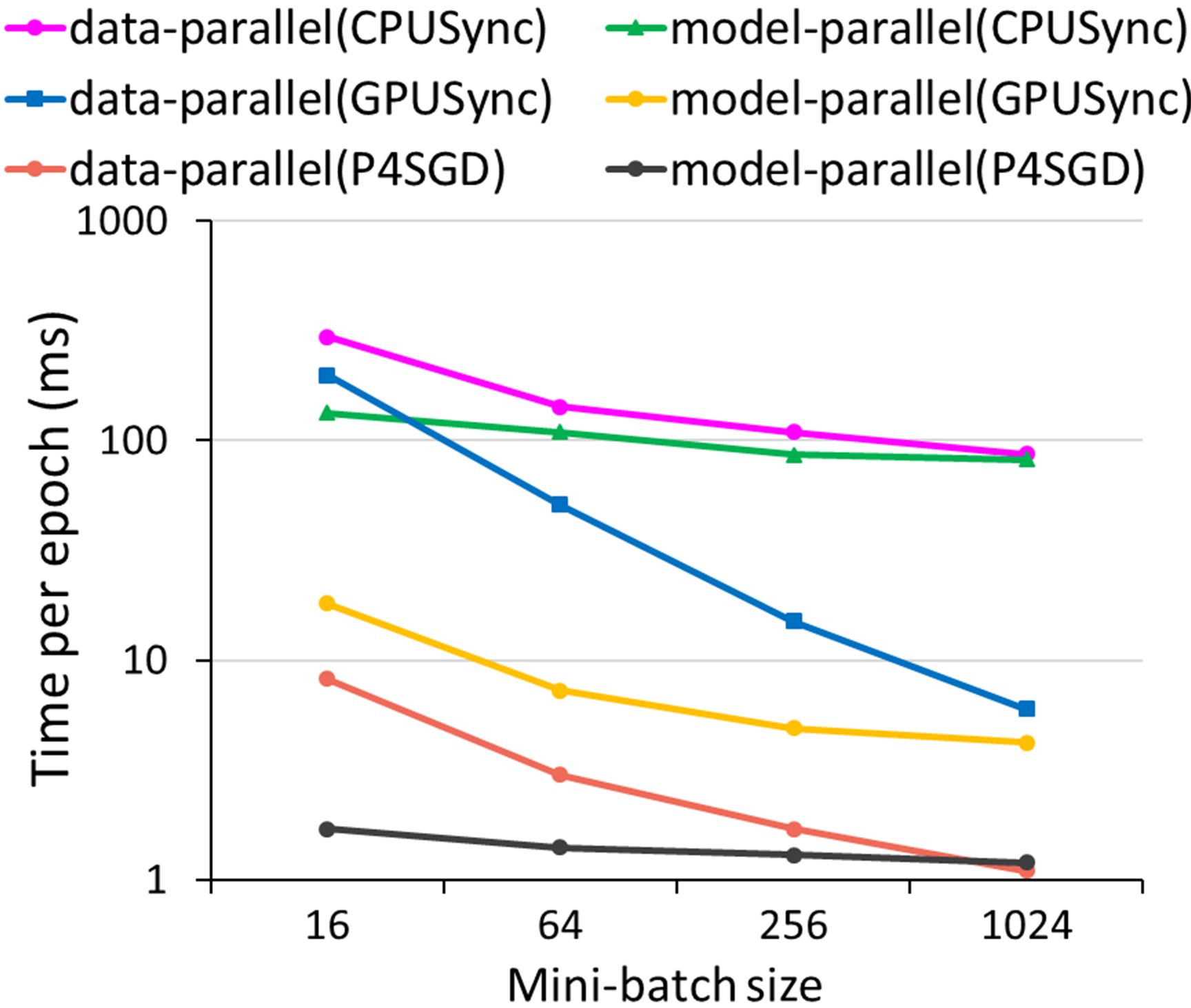} \label{rcv1_data_model_comp}} 
% 	\hfill
% 	\subfloat[$avazu$ (1M)]{\includegraphics[width=3.2in]{figure/avazu_epoch_time_b16.png} \label{avazu_etime_b16}} 	
	%\caption{}
		\vspace{-1ex}	
	\caption{Hardware efficiency comparison between data-parallel and model-parallel, the number of workers is 4.}
	\vspace{-3.5ex}	
% 	\vspace{-3ex}
	\label{data_model_comp} 
\end{figure}

In this section, we compare the time per epoch of data-parallel and model-parallel approaches 
% \marginpar{R2.O4-2}
with other two baselines: ``CPUSync" and ``GPUSync". The number of workers in all experiments is 4. The number of engines of \SystemName{} is 8. Figure~\ref{data_model_comp} illustrates the comparison results under different mini-batch sizes on the dataset $rcv1$ and $amazon\_fasion$. We have three observations.
%\hhj{We compare the convergence of data-parallel and model-parallel with all baseline systems in this subsection. Figure~\ref{data_model_comp} present the results with different mini-batch size on the dataset $real\_sim$ and $rcv1$.}
% First, model parallelism always has significantly smaller elapsed time per epoch than its corresponding data parallelism on the same platform, because model parallelism introduces much less network traffic especially when the mini-batch size is small.
First, model parallelism has significantly smaller elapsed time per epoch than its corresponding data parallelism in most cases on the same platform, because model parallelism introduces less network traffic especially when the mini-batch size is small. Although data parallelism requires larger epoch time when the mini-batch size is large, significantly more epochs are required to reach the same convergence rate, resulting in a smaller overall convergence speedup.
Second, model parallelism has a higher speedup over data parallelism on the same platform when $B$ is smaller, because \SystemName{} introduces hardware pipeline parallelism and P4-switch-FPGA in-switch aggregation with ultra-low latency. For example, when $B$ is 16, model parallelism is $4.8\times$ faster than data parallelism on FPGAs under $amazon\_fasion$, while model and data parallelism has roughly the same elapsed time when $B$ is 1024. Third, model parallelism has a higher speedup over data parallelism on the same platform when the number of features is larger. For example, when $B$ is 16, \SystemName{} is $2\times$ and $4.8\times$ faster than the corresponding data-parallel implementation on FPGAs under datasets $rcv1$ and $amazon\_fasion$, respectively. In the following experiments, we always use model parallelism. 

%\hhj{First, we observe that model-parallel achieves a larger speedup than data-parallel under most situations. The smaller the mini-batch size of training, the larger speedup model-parallel can achieve. For instance, on $amazon\_fasion$, when the mini-batch size is 16, model-parallel is $4.8\times$ faster than data-parallel on \SystemName{}, when the mini-batch size is 1024, model-parallel takes almost the same time as data-parallel on \SystemName{}. Since the communication cost of model-parallel only depends on the mini-batch size, as we discuss in section~\ref{model_parallel_analyse}. 
%Second, the more features the model has, the model-parallel can gain larger speedup. For example, when mini-batch size is 16, model-parallel on \SystemName{} is $2\times$ and $4.8\times$ faster than data-parallel on \SystemName{} on $rcv1$ and $amazon\_fasion$. This is in line with our analysis that the communication time of data-parallel is related to the model size.}

\vspace{-0.5ex}
\subsection{Hardware Efficiency: Throughput}
We examine the hardware efficiency of \SystemName{}, in terms of achievable throughput. First, we examine the effect of different characteristics on \SystemName. Second, we compare \SystemName{} with the baselines on GPUs and CPUs. In the following experiment, by default, we instantiate 8 engines within a worker for best performance. %In addition to the scale-up ability experiment, 

\vspace{-0.5ex}
\subsubsection{Hardware Characteristics of \SystemName{}} %$\\$
We typically run 200 epochs and get the average throughput to analyze the effect of each hardware characteristic. %In the figures of this section, we mark the number of features after the datasets.

\begin{figure}
	\centering
	\includegraphics[width=3.4in]{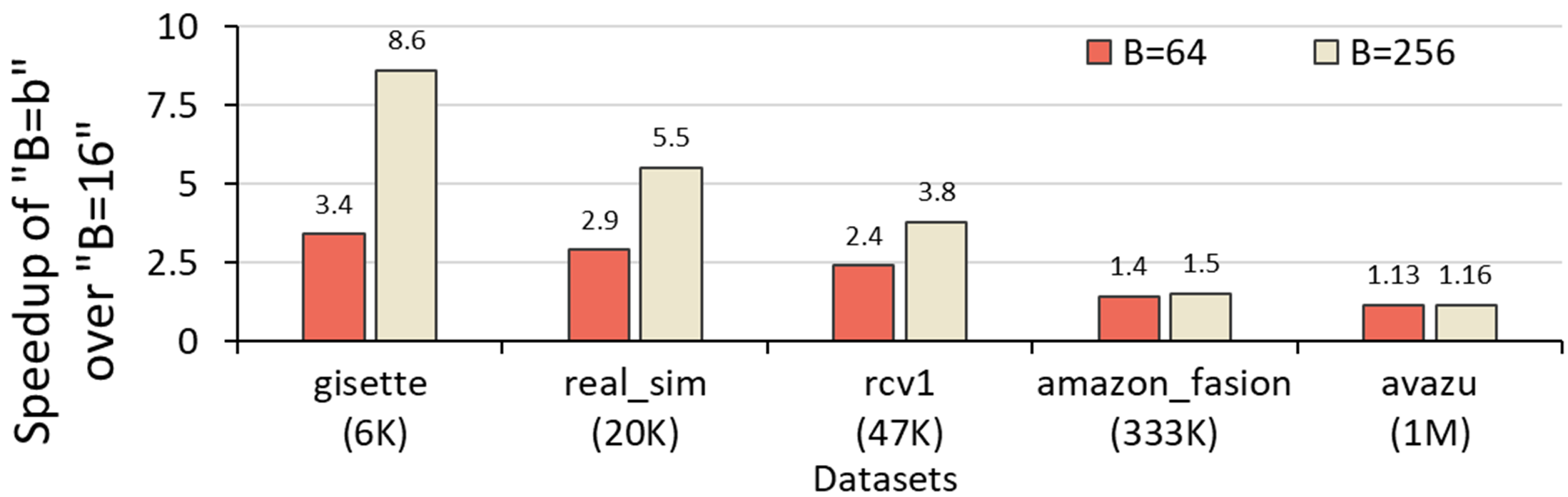} 
	% \caption{}
	\vspace{-1ex}		
	\caption{Effect of mini-batch size, in term of epoch time, on different datasets under the case with 8 workers, each with 8 engines, B: mini-batch size} %,precision = 4 bits
 	\vspace{-2.5ex}
	\label{fpga_mini_batch} 
\end{figure}

\noindent{\bf Effect of Mini-Batch Size. }We examine the effect of mini-batch size on \SystemName{}. We use the implementation of \SystemName{}-8-8. Figure~\ref{fpga_mini_batch} illustrates the \SystemName{}-8-8's speedup of various mini-batch sizes over the case with ``$B$=16", in terms of throughput, on the different datasets. 
We have two observations. First, a larger mini-batch size leads to a higher speedup, since a large mini-batch size allows to overlap forward/backward propagation and communication between micro-batches that belongs to the same mini-batch. %as the mini-batch size increases, the training speed also increases, when the mini-batch size is smaller, the speed of training increases more. 
Second, a larger number of features leads to a smaller speedup when increasing the mini-batch size, because \SystemName{} is able to overlap communication time with computation time that occupies a higher proportion of the total computation time due to a larger number of features, even when $B$ is small. %when the dataset has more features, the acceleration effect caused by the increase of mini-batch size is smaller. 

%The underlying reason for these two observations is that the calculation time of a dataset is stable, the communication time is reduced proportionally according to the increase of mini-batch size, therefore, when the mini-batch size is small, the proportion of the total time occupied by the communication time is large, the acceleration effect is more obvious.
%In the same way, when the feature is small, the calculation takes very little time, so the acceleration effect of the increase in mini-batch size is more obvious.

\begin{figure}
	\centering
	\includegraphics[width=2.3in]{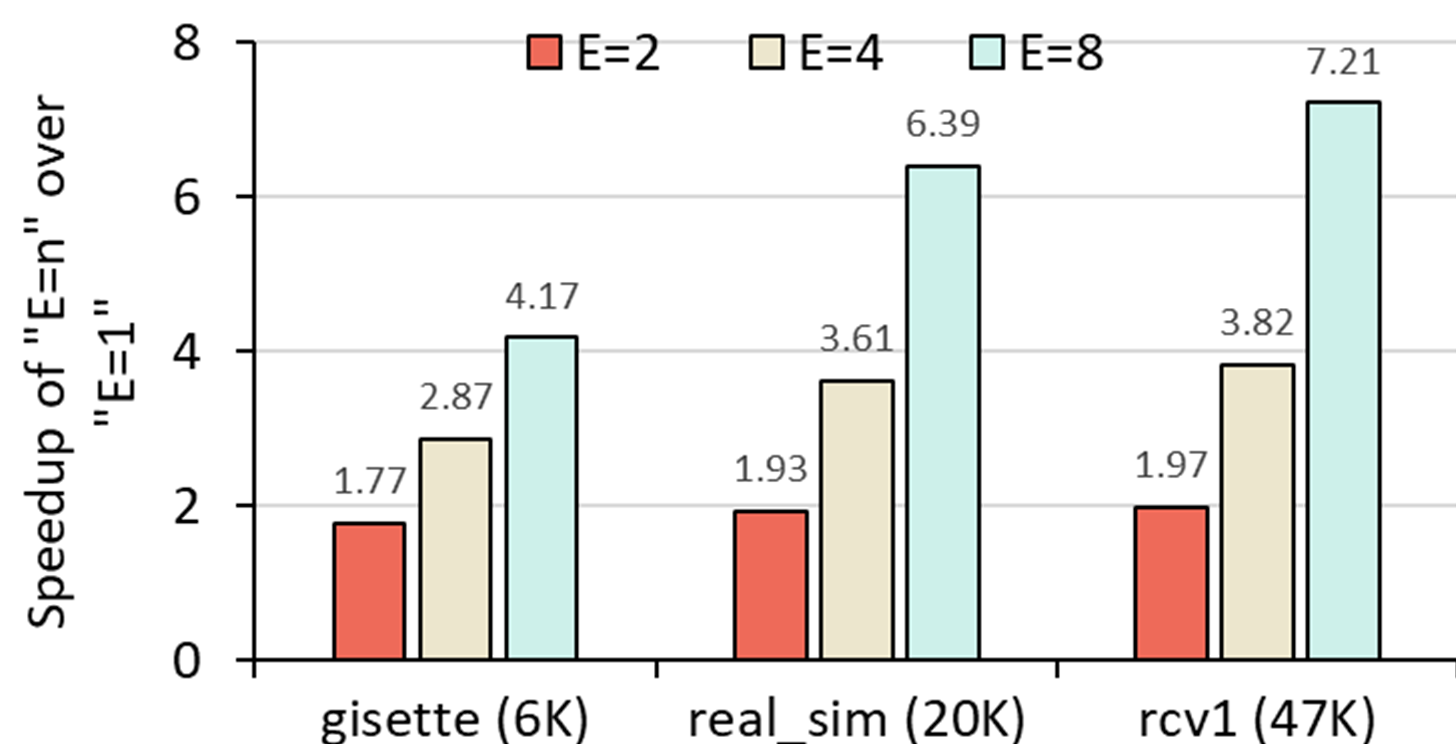} 
	% \caption{}
	\vspace{-1ex}		
	\caption{Scale-up ability, in term of epoch time, under the case with 1 worker, E: number of engines, mini-batch size = 64, and precision = 4 bits} 
	\vspace{-3ex}
	\label{fpga_engines} 
\end{figure}

\noindent{\bf Scale-up Ability. } We examine the scale-up ability of \SystemName{} that instantiates multiple engines for model-parallel training on GLMs. Figure~\ref{fpga_engines} shows the throughput ratio of the cases with a various number of engines over the case with one engine on the datasets $gisette$, $real\_sim$, and $rcv1$. We have two observations. First, a larger number of engines leads to higher throughput due to fewer training tasks per engine within a worker. Second, a larger number of features leads to a higher throughput improvement when increasing the number of engines, because a larger number of features leads to a higher proportion of computation time to overall training time, where computation time can be linearly reduced by introducing more engines.   %We train \SystemName{}-1-1, \SystemName{}-1-2, \SystemName{}-1-4 and \SystemName{}-1-8 on $gisette, real\_sim, rcv1$ datasets. Figure~\ref{fpga_engines} presents the result.  We observe that the performance of \SystemName{} improves is sub-linear as the number of engines increases on $rcv1$, but the performance change on $gisette$ is relatively small. Because when the engines increase, the amount of calculation for training decreases proportionally, but the amount of data transmitted does not change. We conclude that the larger the model, the larger the mini-batch size, the \SystemName{} can more significantly reduce the training time.

% {\bf Local Server and Remote Server.} We implement \SystemName{}-1 with two methods. One method is to implement the worker and server on an FPGA($local$ $server$), the other method is to implement worker and server on two FPGA respectively($remote$ $server$). We examine these two methods with various engines and batch size on $real-sim$ and $rcv1$ datasets. Figure\ref{server_b_gi} and Figure\ref{server_b_rc} shows the effect of batch size. We observe that the training time of local server is roughly stable for different batch size, since local server have no network communication, and the \SystemName{} commendably overlap computation, regardless of batch size. In contrast, the performance of remote server is affected by batch size and model size, the less network traffic relative to calculated amount, the better the overall performance. But the performance of remote server is 
% similar to local server when batch size is 256 on $rcv1$. We conclude that although the performance of remote server is not as good as local server, when the model size is tens of thousands, and the batch size is 256, the performance of remote server is almost the same as that of local server. 

\begin{figure}
	\centering
% 	\subfloat[FPGA-only]{\includegraphics[width=3.2in]{figure/fpga_scale_out.png} \label{fpga_scaleout}} 
% 	\hfill
	\includegraphics[width=2.5in]{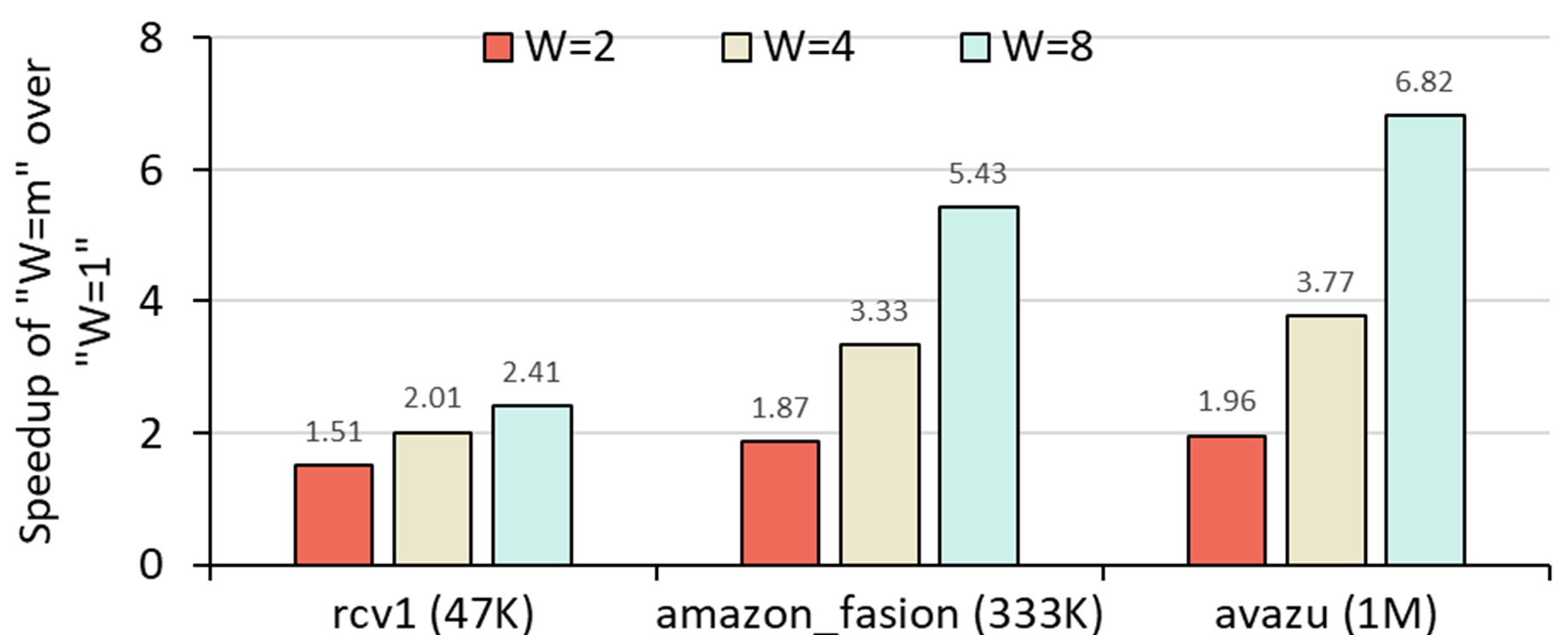}
	%\caption{}
		\vspace{-0.5ex}	
	\caption{Scale-out ability of \SystemName{} in term of epoch time with 8 engines, W: number of workers, mini-batch size = 16, and precision = 4 bits}
	\vspace{-3.5ex}	
% 	\vspace{-3ex}
	\label{dltraining_scale_out} 
\end{figure}

 \noindent{\bf Scale-out Ability.} We examine the scale-out ability of \SystemName{} that employs multiple workers for model-parallel training on GLMs. Figure~\ref{dltraining_scale_out} shows the throughput ratio of the cases with a various number of workers over the case with one worker. We have three observations. First, a larger number of workers leads to a higher throughput due to fewer training tasks per worker. Second, a larger number of features leads to a higher throughput improvement when increasing the number of workers, because a larger number of features leads to a higher proportion of computation time to overall training time, which can more easily amortize the negative effect of communication time.
%  \marginpar{R1.O1-4} 
Third, when the number of features reaches 1 million, the throughput closely increases linearly with an increasing number of machines. It indicates that \SystemName{} achieves a strong scale-out ability when the number of features is large enough ( $> 1M$). 

\hhj{}

\subsubsection{Scalability Comparison with CPU/GPU Baselines} %$\\$

\begin{figure}[t]
	\centering
	\subfloat[$rcv1$ (47K), B=16 ]{\includegraphics[width=1.55in]{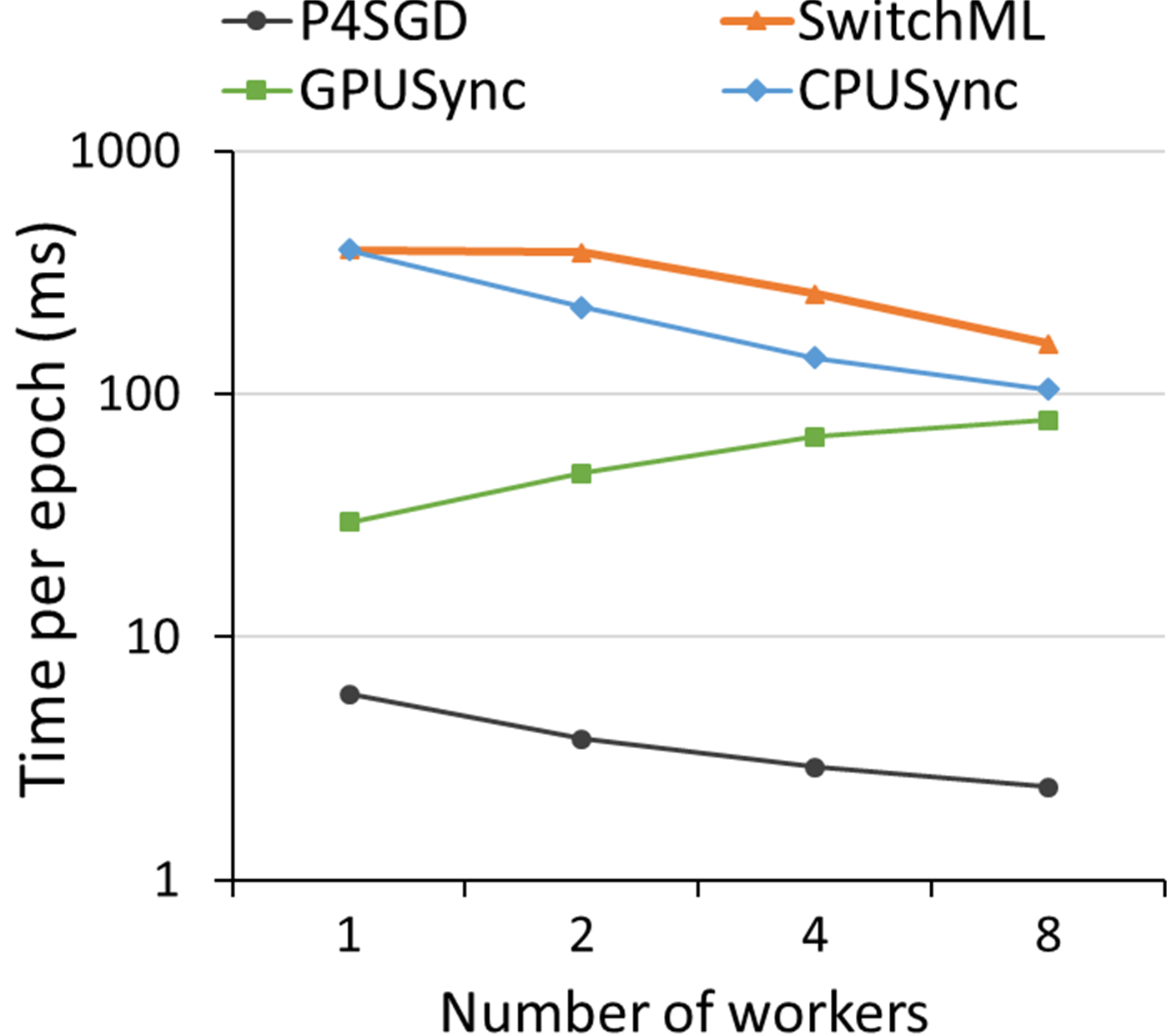} \label{rcv1_etime_b16}} 
	\subfloat[$amazon\_fashion$ (333K), B=16]{\includegraphics[width=1.6in]{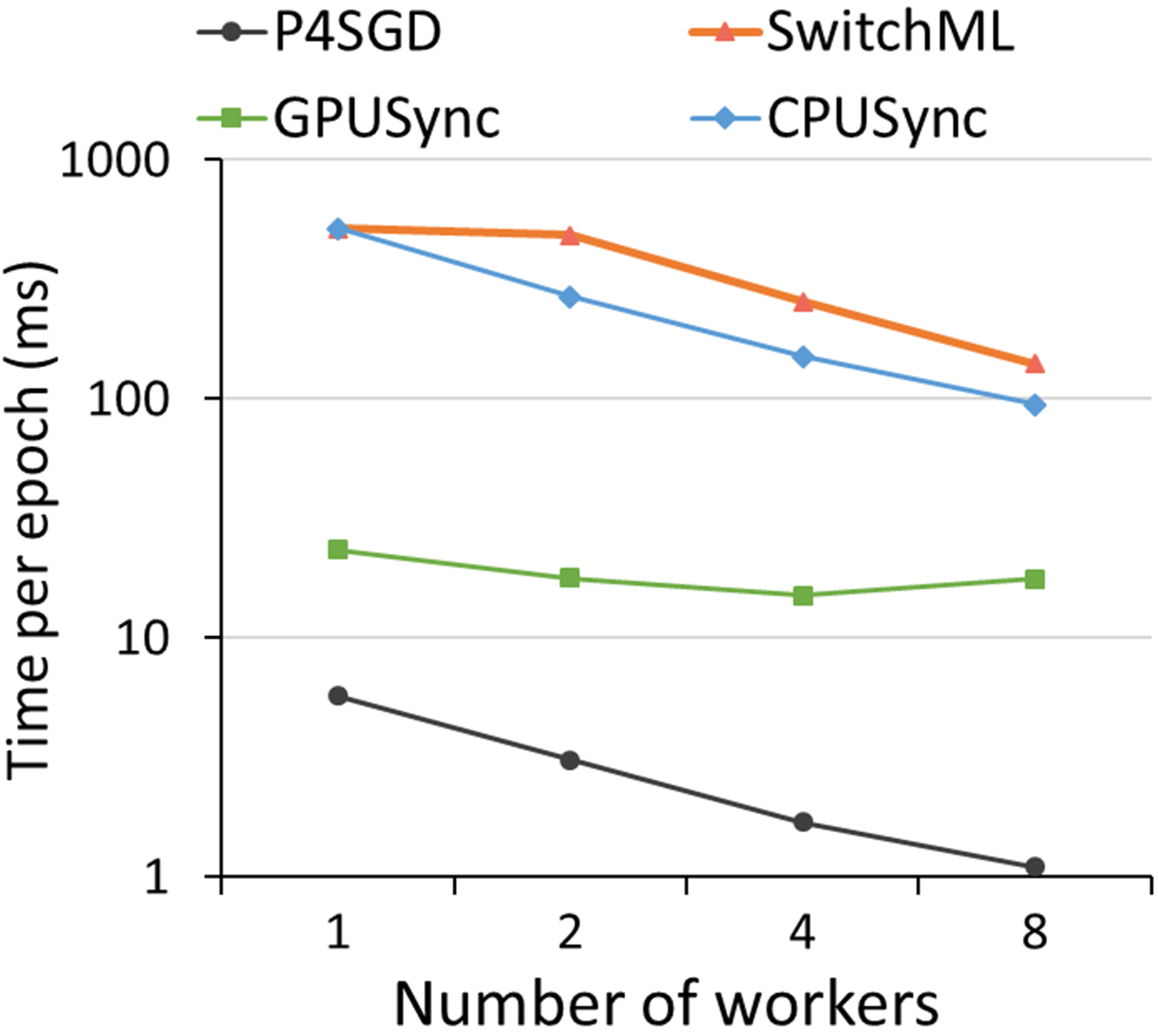} \label{amazon_etime_b16}} 
	\hfill
	\subfloat[$rcv1$ (47K), B=256]{\includegraphics[width=1.55in]{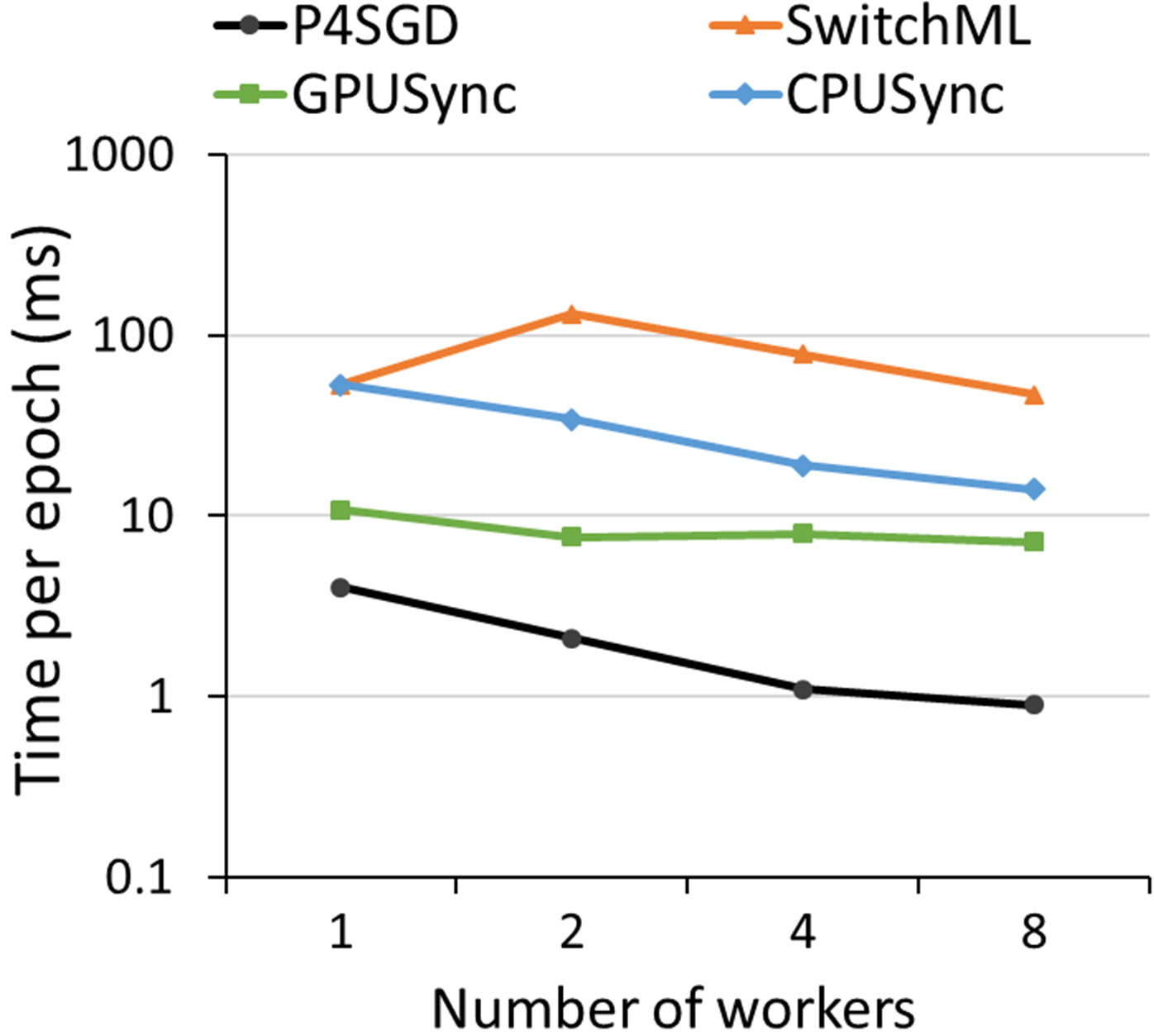} \label{rcv1_etime_b256}} 
	\subfloat[$amazon\_fashion$ (333K), B=256]{\includegraphics[width=1.6in]{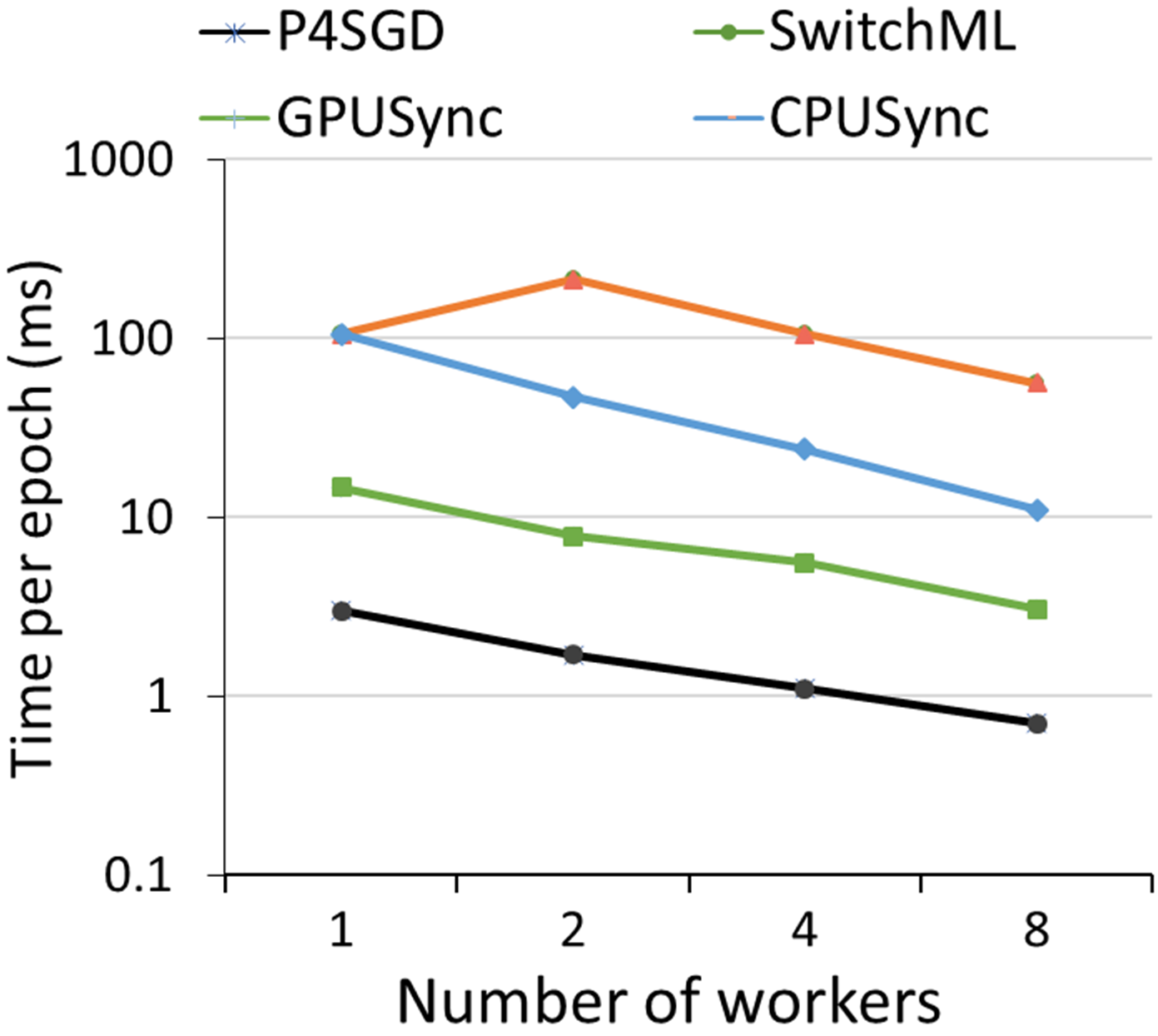} \label{amazon_etime_b256}}
	%\caption{}
		\vspace{-1ex}	
	\caption{Hardware efficiency comparison between \SystemName{} and other baselines}
	\vspace{-3.5ex}	
% 	\vspace{-3ex}
	\label{epoch_time} 
\end{figure}

% \hhj{\marginpar{R1.O3-3}
We compare scalability, in terms of time per epoch, of \SystemName{} with other three baselines: ``SwitchML``, ``CPUSync", and ``GPUSync". Figure~\ref{epoch_time} illustrates the average epoch time with the various number of workers and mini-batch size. We have four observations. 

First, ``\SystemName{}" is significantly faster than the other three counterparts and has the highest scalability. This is due to its latency-centric in-switch aggregation protocol and hardware pipeline parallelism to maximize the overlap between communication and forward/backward propagation. 

Second, ``GPUSync" fails to scale out when $B$ is relatively small, because 1) it cannot fully utilize the GPU computing power due to severe kernel invocation overhead (innovating three kernels per iteration), and 2) more distributed GPUs exacerbate kernel invocation overhead and thus amortize the benefit of reduced computation time of the CUDA kernel \emph{cublasSgemm}, especially when the dimension is relatively small, as shown in Figure~\ref{rcv1_etime_b16}. High-performance intra-node NVLink slightly relieves the communication overhead, as the communication stage only accounts for roughly 20\% of the total training time when using RDMA-GPUDirect-powered NCCL. 

Third, ``CPUSync" can relatively easily scale out, because computation time dominates the overall training time on distributed CPUs, and communication time is negligible. Therefore, when the number of workers increases, the overall training time can drop quickly.

Forth, ``SwitchML" is slower than ``CPUSync", and its scale out ability is also worse than that of ``CPUSync". The mainly reason for this is that ``SwitchML" has the highest aggregation latency due to its shadow copy mechanism that delays the acknowledgement of received aggregation packets, as shown in Figure~\ref{fig_control_path_latency}. Actually, ``SwitchML" adopts the shadow copy mechanism to greatly increase the throughput of in-network aggregation, rather than decreasing latency.

% \begin{figure*}
% 	\centering
% 	\subfloat[Gisette, precision = 4]{\includegraphics[width=2.3in]{figure/scale_gisette_4bit.png} \label{scale_gisette_b4}} 
% 	\subfloat[Real-sim, precision = 4]{\includegraphics[width=2.3in]{figure/scale_real_sim_4bit.png} \label{scale_readsim_b4}}	
% 	\subfloat[Rcv1, precision = 4]{\includegraphics[width=2.3in]{figure/scale_rcv1_4bit.png} \label{scale_rcv1_b4}}	
% 	\hfill
% 	\subfloat[Gisette, precision = 8]{\includegraphics[width=2.3in]{figure/scale_gisette_8bit.png} \label{scale_gisette_b8}} 
% 	\subfloat[Real-sim, precision = 8]{\includegraphics[width=2.3in]{figure/scale_real_sim_8bit.png} \label{scale_readsim_b8}}	
% 	\subfloat[Rcv1, precision = 8]{\includegraphics[width=2.3in]{figure/scale_rcv1_8bit.png} \label{scale_rcv1_b8}}
% 	%\caption{}
% 	%\vspace{-1ex}	
% 	\caption{local server and remote server comparison. precision = 4}
% 	\vspace{-3ex}
% 	\label{fig_mm} 
% \end{figure*}  

\vspace{-1.5ex}
\subsection{Statistical Efficiency: Loss vs. Epochs}
\begin{figure}[t]
	\centering
	\subfloat[$rcv1$ (47K)]{\includegraphics[width=1.5in]{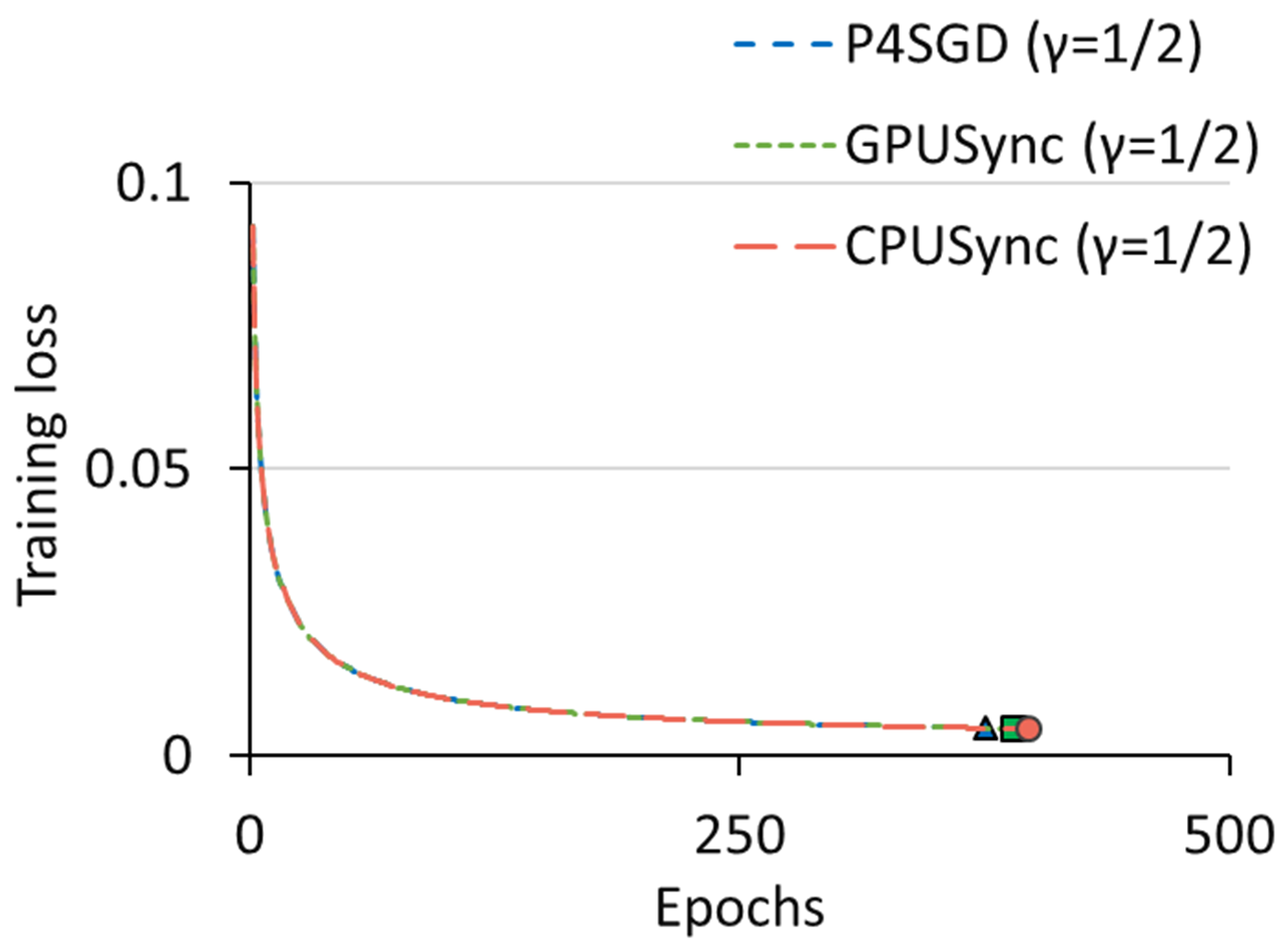} \label{network_fpga}} 
	\subfloat[$avazu$ (1M)]{\includegraphics[width=1.5in]{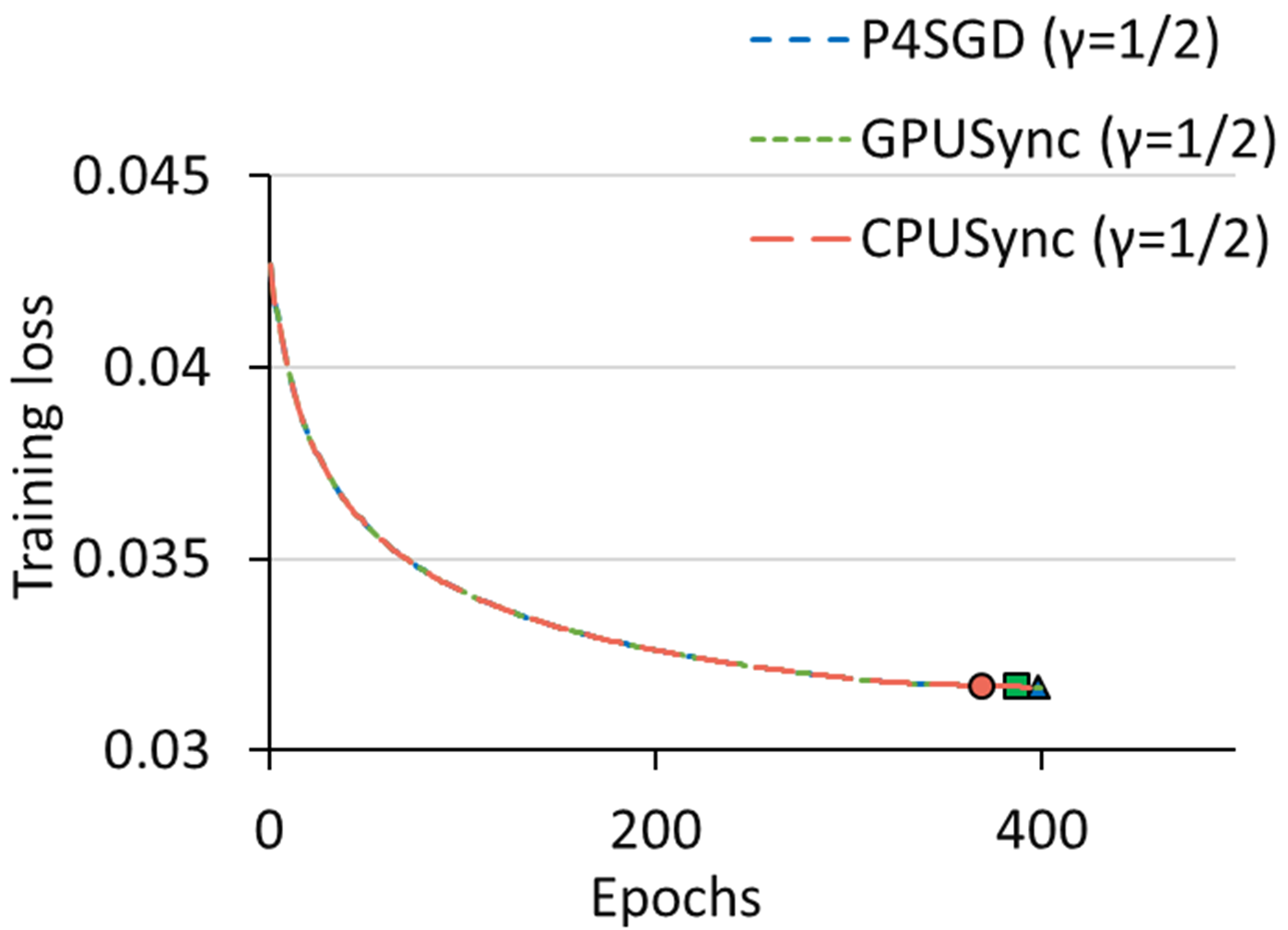} \label{network_p4}}
	\vspace{-1ex}	
	\caption{Statistical efficiency: training loss vs. epoch, $\gamma$: learning rate, precision = 4 bits} 
 	\vspace{-3.5ex}
	\label{network_fig} 
\end{figure}

We compare the statistical efficiency of \SystemName{} with ``CPUSync", and ``GPUSync" on the datasets $rcv1$ and $avazu$. %We use part of the samples of the $avazu$ dataset for training because the dataset is too large to fit the external memory on FPGA. 
Figure~\ref{network_fig} shows the convergence trend under an increasing number of epochs. The batch sizes for all approaches are 64. We observe that all the methods require the same number of epochs to converge to roughly the same loss, because they are all synchronous. %In the $avazu$ dataset, the \SystemName{} has a faster convergence speed because of the smaller mini-batch size.

\vspace{-1.5ex}
\subsection{End-to-End Comparison: Loss vs. Time} \label{endtoend}
\begin{figure}[t]
	\centering
	\subfloat[$rcv1$ (47K)]{\includegraphics[width=1.7in]{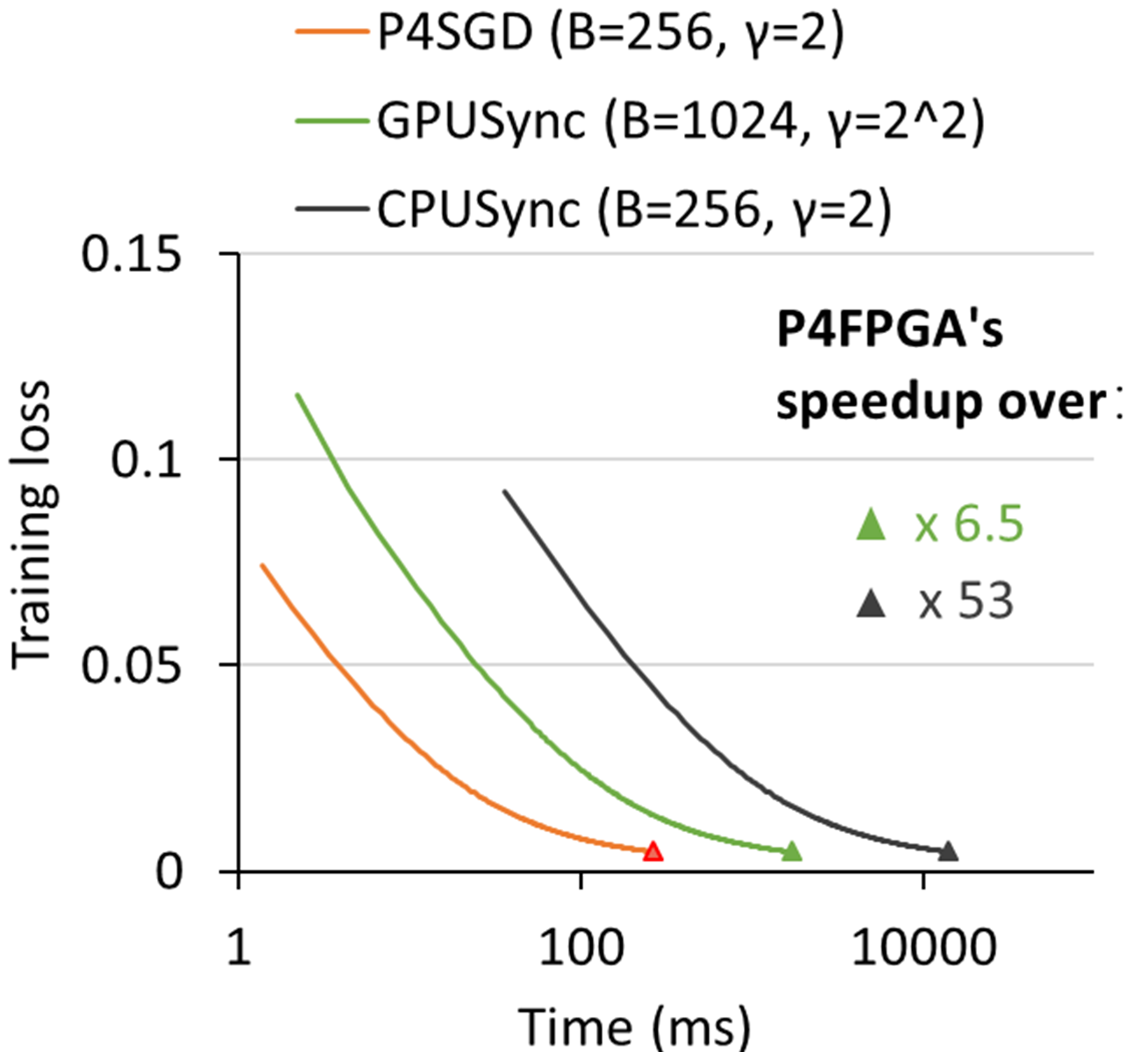} \label{lossrcv1}} 
	\subfloat[$avazu$ (1M)]{\includegraphics[width=1.7in]{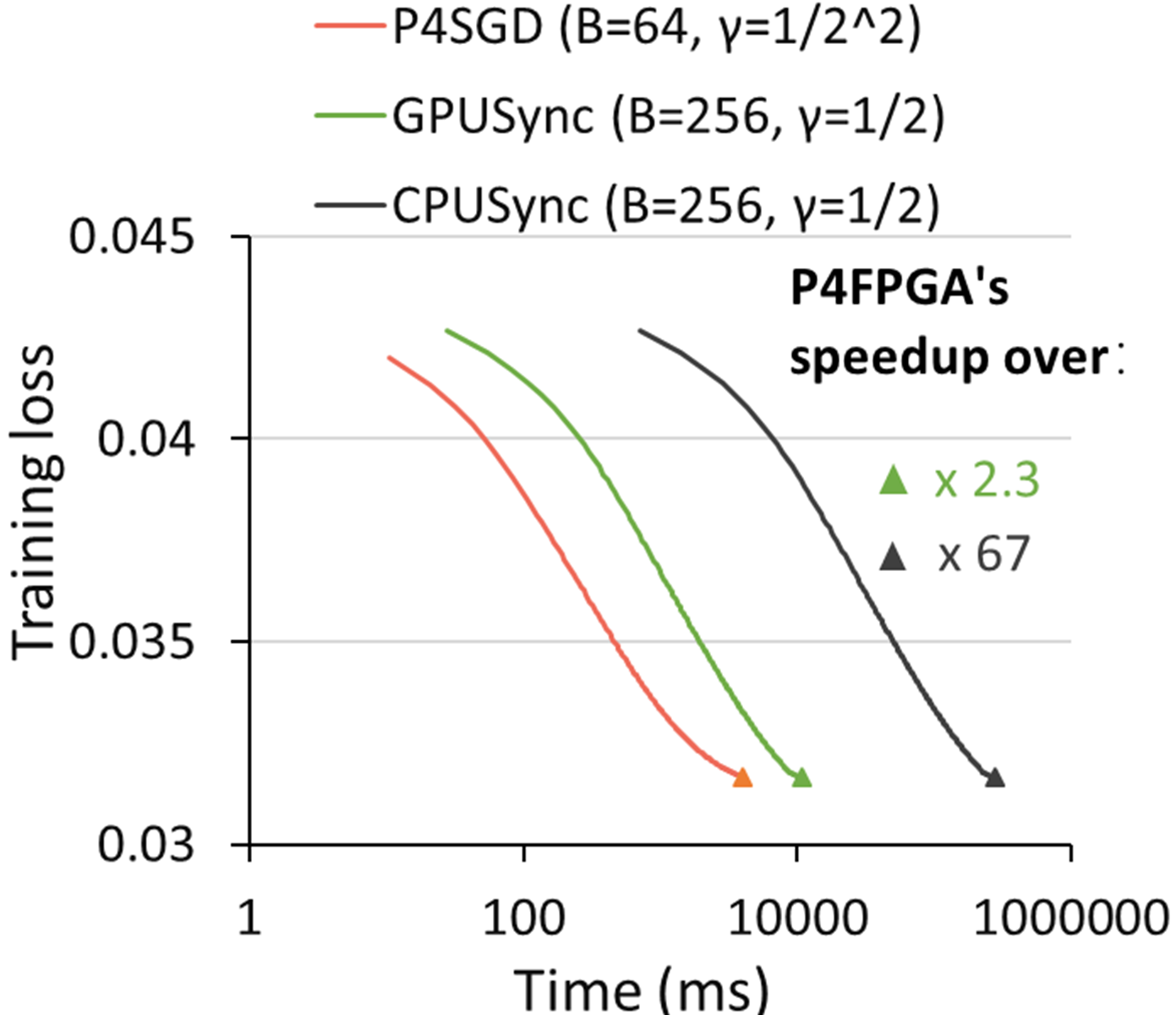} \label{lossavazu}} 
	%\caption{}
	\vspace{-1ex}	
	\caption{End-to-end convergence comparison: training loss vs. time, B: mini-batch size, $\gamma$: learning rate, precision = 4 bits}
	\vspace{-2ex}	
% 	\vspace{-3ex}
	\label{loss_time} 
\end{figure}

\vspace{-0.5ex}
We compare the end-to-end performance of \SystemName{} with ``CPUSync" and ``GPUSync", which use the configurations with the shortest convergence time. 
% For example, in the $rcv1$ dataset, ``GPUSync" uses 1 worker, ``FPGA-only" uses 2 workers, \SystemName{} and ``CPUSync" uses 8 workers.
Figure~\ref{loss_time} shows the end-to-end convergence comparison, in terms of loss vs. time, on the datasets $rcv1$ and $avazu$. We have two observations. First, \SystemName{} is able to converge up to 6.5X faster than ``GPUSync", indicating great potential on in-network aggregation and micro-batch model-parallel training that enable efficient model-parallel training on distributed FPGAs. ``GPUSync" converges relatively slowly, due to 1) long communication overhead per iteration, and 2) no overlap between forward/backward propagation and communication.  %\wzk{The reason why GPUSync converges relatively slowly is that GPU needs a large mini-batch size, e.g., 1024, to fully utilize GPU's massive computing power and thus achieves higher hardware efficiency than \SystemName{}; however, the large mini-batch size reduces the number of iterations and thus is harmful to the end-to-end convergence rate. Actually, GPUSync can converge significantly faster with a mini-batch size of 256 than with a mini-batch size of 1024 on the dataset $avazu$, as shown in Figure~\ref{lossavazu}, even though a mini-batch of 1024 can fully utilize GPU's computing power and thus has higher hardware efficiency (i.e., time per epoch). Therefore, GPUSync chooses a mini-batch size of 256 rather than 1024, as shown in Figure~\ref{lossavazu}. }
Second, \SystemName{} can converge up to 67X faster than ``CPUSync", because ``CPUSync" suffers from low compute power and long collective operation latency. 
%{\bf Loss vs. Time.} Figure~\ref{loss_time} shows the comparison result. We have three observations. First, \SystemName{} uses the synchronous SGD algorithm with a precision of 4, and its loss is almost the same as the loss calculated by the full-precision algorithm. Therefore, a 4-bit precision is sufficient for linear model SGD training. Second, the CPU is not suitable for high-parallel calculations, so in the training process, its speed is many times slower than FPGA and GPU. Third, due to the reduction of communication time, the total training time has been significantly reduced.

\subsection{Energy Consumption}
\vspace{-0.5ex}
% \marginpar{R2.O4-4}
In addition to hardware efficiency and performance, energy consumption has become a significant consideration in recent years. \SystemName{} has a significant advantage in terms of energy consumption.

% \marginpar{R2.O4}\hhj{
To measure the power consumption of the Xilinx Alveo U280, we use the Alveo Card Management Solution Subsystem (CMS Subsystem). The CMS firmware is responsible for gathering the U280's voltage, current, power consumption, and other related information from the satellite controller device.\cite{cms}
% To measure the power consumption of the Xilinx Alveo U280, we use the Xilinx Power Estimator (XPE) in Vivado SDK version 2020.2. The XPE estimates the power consumption of \SystemName{}. 
The power consumption of the ``CPUSync" is measured using lm\_sensors \cite{lm-sensor}, an open-source application that provides tools and drivers for monitoring temperatures, voltage, and power. The power consumption of the ``GPUSync" is measured using the NVIDIA System Management Interface (nvidia-smi) \cite{nvidia-smi}, which can manage and monitor NVIDIA GPU devices.

We evaluate the power consumption of \SystemName{} against "CPUSync" and "GPUSync" in the end-to-end experiments in Section~\ref{endtoend}. Table~\ref{energy_consumption} presents the energy consumption comparison, which does not include the power consumption of the host system. The results show that \SystemName{} is up to 11X more energy-efficient than ``GPUSync" and 50X more energy-efficient than ``CPUSync", demonstrating the significant advantage of \SystemName{} in terms of energy consumption.

\begin{table} [t]
	\centering
	%\begin{spacing}{0.3}
		\begin{scriptsize}
	\vspace{-0.5ex}
	\caption{Energe consumption, the number of workers for all methods is 8.}% under \SystemName{}
	\label{energy_consumption}
	\vspace{-1ex}
	\begin{tabular}{|c||c|c|c|c|c|}
		\hline
		{\bf Method} &  {\bf Dataset} &  {\bf Time(s)} & {\bf Total Power(W)} &  {\bf Energy(J)} \\
		\hline
		\hline
		{\bf P4SGD} & rcv1 & 0.27 & 528 & 143 \\ 
		& avazu & 4.12 & 528 & 2175 \\ 		
		\hline
		{\bf GPUSync} & rcv1 & 1.76 & 920  & 1619\\ 
         & avazu & 10.9 & 920 & 10028\\ 		
		\hline		
		{\bf CPUSync} & rcv1 & 14.4 & 496 & 7142  \\ 
		 & avazu & 128.25 & 496  & 63612  \\ 		
		\hline
	\end{tabular}
	\vspace{-2ex}

		\end{scriptsize}
\end{table}

\vspace{-1.5ex}
\section{Related work}

\noindent\textbf{Parallelism in Distributed ML. }ColumnSGD~\cite{ColumnSGD2020icde} proposes the distributed SGD training system using model parallelism on multiple CPUs. 
% \hhj{\marginpar{R1.O3-3}
Its iteration time remains almost unchanged as the number of machines increases. Even worse, ColumnSGD increases the per-iteration time by 1.3 times, when the number of machines increases from 10 to 40. Megatron\-LM~\cite{shoeybi2020megatronlm} efficiently trains multi-billion language models with model parallelism between GPUs within a node and with data parallelism between nodes. %~\cite{prasad2015large} implements a large-scale predictive analytics workflow which allows row-oriented and column-oriented data access. %, DMac~\cite{yu2015exploiting} proposes a matrix computation system for distributed ML including both row-partitioning and column-partitioning. by cleverly partitions weights of consecutive linear layers by column and by row, therefore reducing communication frequency.  
Alpa~\cite{zheng2022alpa} automates inter- and intra-operator parallelism for training a \emph{large} DL model that cannot fit in a GPU. Alpa recommends intra-operator parallelism, due to its high communication overhead, to accelerators that are connected with high-bandwidth communication link like NVLink, and recommends inter-operator parallelism to distributed accelerators that are connected by relatively low-bandwidth network. GSPMD~\cite{xu2021gspmd} auto-completes the sharding on every tensor based on user annotations, the user can combine data-, model-, and pipeline- parallelism, etc. However, GSPMD does not partition individual operators or tensors in pipeline parallelism, but partitions the training graph into multiple stages that run on different devices. However, \SystemName{} strives to efficiently, e.g., in strong scaling, leverage intra-operator parallelism (model parallelism in our paper) to train a fully-connected layer (i.e., GLM model) between distributed accelerators such as FPGA with the help of a programmable switch. %\hhj{Alpa~\cite{zheng2022alpa} trains dense DL models with automatic 3D parallelism, whereas P4FPGA targets much more sparse GLM models tensor parallelism and fine-grained pipeline parallelism. We believe our parallelism strategy represents an optimal configuration in the Alpa search space, and we carefully optimized the model parallelism through pipelining and latency hiding. Thus, we expect our results to be on par or better than Alpa's. Additionally, our contribution includes running GLMs efficiently in a novel FPGA environment whereas Alpa targets traditional GPU clusters.} 
% Alpa trains large deep learning(DL) models across two levels of parallelism. Contrary to our work, Alpa chooses model parallelism across multiple GPUs within a node, and pipeline parallelism across multiple nodes. This is because DL models have much more weights than activations and data in one mini-batch, and GLMs are the opposite, so we choose a different parallel strategy. At the same time, the communication amount of GLM is much more than that of DL relative to the amount of calculation. Therefore, we chose P4+FPGA to speed up allreduce and reduce communication time. However, there is no Alpa source code for the current paper on the Internet, so we added a discussion about alpa in section VI. 
%\hhj{However, \SystemName{} targets GLM models and partitions data into micro-batch in one device.}

\noindent\textbf{Pipeline Parallelism in Distributed ML. }Previous pipeline parallelism, e.g., GPipe~\cite{gpipe_NEURIPS2019}, HetPipe~\cite{park2020hetpipe}, TeraPipe~\cite{terapipe_arxiv21} assigns different layers of a model to different GPUs to train a large model, and thus is orthogonal to \SystemName, which targets model parallelism. %, so at the beginning of each iteration, only the GPU with the first layer can start working first, and other GPUs need to wait for the first micro-batch calculation to complete.
% The model parallelism like Megatron-LM~\cite{} divides the model of the same layer to different GPU to avoid this waste, but it only supports model parallelism within the node. 
%We divide the models of the same layer into different nodes, and divide a mini-batch of samples into smaller micro-batches to take full advantage of the performance of each worker.

\noindent\textbf{FPGA-accelerated ML systems.} \SystemName{} is closest to MLWeaving~\cite{mlweaving2019vldb}, which is a one-size-fits-all system for any-precision learning.
% \marginpar{R2.O4-3}
The hardware efficiency of MLWeaving is comparable to that of \SystemName{} using 1 engine and 1 worker. Prior works~\cite{boutros2018embracing, cadambi2009massively, sharma2016high, caulfield2016cloud, chiu2018flexibility, chung2018serving, de2017understanding, fowers2018configurable, kara2017fpga, li2019rnn, mahajan2018rdbms, mahajan2016tabla, moreau2015snnap, dass2020distributed} exploit FPGAs 
to achieve high performance, but not for model-parallel training. Brainwave~\cite{chung2018serving} accelerates DNN inference via model parallelism on multiple FPGAs. Previous work~\cite{dass2020distributed} presents a multiple-FPGA system for accelerating distributed SVM training, but all the FPGAs are on one host, where they communicate with each other through PCIe. Therefore, its scalability is limited in a single host.

%\textbf{SGD in ML/DL System.} Most of today's most advanced training systems use data parallelism to partition the data when implementing SGD. TensorFlow\cite{abadi2016tensorflow} is a widely used machine learning system that operates at large scale and it can not split a model into pieces and store them on different servers. Petuum\cite{xing2015petuum} is a new platform for distributed machine learning and  it contains model parallelism. Spark\cite{zaharia2010spark,meng2016mllib} is a popular platform for machine learning, MLlib is the library of Spark. In MLlib, the model is generally stored on the master or parameter servers, workers store a part of the training data and pull the model when training SGD. MXNet\cite{chen2015mxnet} is also a distributed training system, in particular, it provides library for deep neural networks.  

\noindent\textbf{In-network Aggregation. } Current research ATP~\cite{lao2021atp}, SwitchML~\cite{sapio2019scaling} and~\cite{luo2018parameter} leverages programmable network switch to perform gradient aggregation in data parallel training. Also, they adopt shadow copy mechanism to optimize for throughput. In contrast, \SystemName{} optimizes for ultra low in-network aggregation latency to benefit our model-parallel training between distributed FPGAs. 
SHARP~\cite{graham2016scalable} offloads collective operation processing to the InfiniBand network. SHARP approaches a hierarchical aggregation tree architecture to enable general aggregation operations on hundreds of hosts.  Herring~\cite{thangakrishnan2020herring} is a scalable distributed data-parallel training library-based parameter server that adopts a balanced fusion buffer to solve the problem of unbalanced data sent and received by servers. Herring targets data-parallel distributed training mainly via increasing the throughput of aggregation operations, while \SystemName{} targets model-parallel distributed training mainly via reducing the latency of aggregation operations.  %However, P4FPGA offloads the aggregation operation to the Ethernet, and approaches a new protocol optimized for short messages in GLMs training which achieves shorter latency.

\vspace{-2.5ex}
\section{Conclusion}
We propose \SystemName{}, a model-parallel distributed training system that allows strong scaling when training GLMs. \SystemName{} adopts micro-batch hardware pipeline-parallel training to overlap forward/backward propagation and communication within a worker. At the same time, we propose a latency-centric in-switch aggregation protocol to lower communication overhead between distributed FPGAs. We prototype \SystemName{} on multiple FPGAs and a P4 switch. The experimental result shows that \SystemName{} is able to converge up to 9.3x faster than its GPU counterpart. We will make \SystemName{} open-sourced to benefit the community. %We performed an analytic comparison between data parallelism, vanilla model parallelism and \SystemName{} model parallelism. We further provided a network protocol to ensure reliable data transmission between the workers and the server. We have also implemented \SystemName{} and baselines on a cluster with 8 machines and 2 network switches and conducted experimental evaluations to prove its effectiveness.

%\wzk{\noindent {\bf Future Work. } The common wisdom is that model parallelism only works between accelerators within a compute node. In this paper, we revisit this challenge by achieving linear scalability between distributed accelerators. In particular, we revisit the training of generalized linear models (GLMs). The GLM training is similar to the training of a fully-connected layer that has so low compute density that the reduced compute time from introducing more accelerators throughput is not able to amortize the communication overhead.}

\noindent {\bf Limitation and Future Work. }The main limitation is that \SystemName{} implements the model on FPGA's on-chip memory, so the model size is limited due to the limited on-chip memory size. Our current implementation supports parameterizable model size, but up to 2M. However, we can easily generalize \SystemName{} to support a large model size by storing the model in external memory, e.g., HBM. As such, the implementation for a large model needs more memory bandwidth to achieve line-rate hardware processing. Fortunately, the current \SystemName{} only uses 25\% HBM's memory bandwidth, leaving the majority of memory bandwidth for engines to access a large model in external memory. We believe \SystemName{} would not lose any processing speed. We leave the related implementation in future work.

%\hhj{Another limitation of \SystemName{} is that the FPGA must be reconfigured when the number of engines in the worker is changed, which takes a few minutes. Therefore, we usually use the maximum number of engines that can be used under the FPGA resource limitation. In future implementations, we may change the number of engines to an input variable so that the system does not need to be reconfigured. }

Another future work is to generalize \SystemName{} to DNN training using an FPGA-GPU co-processing approach. In particular, we train compute-intensive layers, e.g., convolutional, on GPUs via data parallelism, since these layers have a high computation amount per weight, while we train the fully-connected layers on \SystemName{} via model parallelism since these layers may have a big model size but have a low computation amount per weight. 

% \marginpar{R2.O3-2}
The third future work is to run the \SystemName{} at a frequency of 225MHz, enabling it to process 512-bit data from a single HBM channel, thereby doubling the number of engines supported by the current design.
%\hhj{The third future work is scaling the P4FPGA to more than one switch. At present, our in-network aggregation protocol only supports a single switch and $n$ workers while the $agg\_bm$ (in \autoref{alg_switch_unreliable}) is $n$ bits. However, we can make it support second-level switch by simply modifying the aggregation protocol, such as adding a $n$ bits $switch\_bm$ field to indicate the first-level switch's position at the second-level switch. Then the system can support up to $n^2$ workers. Although two-level switches double the latency of the $AllReduce$ operation, in the training of the billions of model size which two-level switches is necessary, the impact of the scalability is very limited.}

\noindent {\bf Acknowledgements. }
The work is supported by the following grants: the National Key R\&D Program of China (Grant No. 2022ZD0119301), the Program of Zhejiang Province Science and Technology (2022C01044), the Fundamental Research Funds for the Central Universities 226-2022-00151, Starry Night Science Fund of Zhejiang University Shanghai Institute for Advanced Study (SN-ZJU-SIAS-0010), Key Laboratory for Corneal Diseases Research of Zhejiang Province, a research grant from Alibaba Group through Alibaba Innovative Research (AIR) Program.

% \vspace{-2.5ex}
% \input{context/artifact.tex}

\bibliographystyle{IEEEtran}
\bibliography{references}

\begin{IEEEbiography}[{\includegraphics[width=1in,height=1.25in,clip,keepaspectratio]{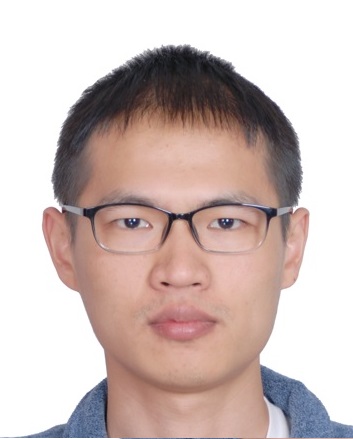}}]{Hongjing Huang}
is currently a Eng.D. student at Zhejiang University, China. Prior to that, he received his master's and bachelor's degree from Zhejiang University. His research interests include distributed machine learning, SmartNIC, etc.
\end{IEEEbiography}

\begin{IEEEbiography}[{\includegraphics[width=1in,height=1.25in,clip,keepaspectratio]{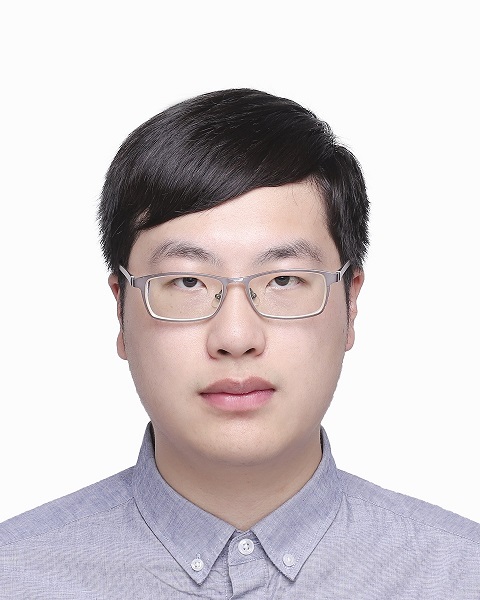}}]{Yingtao Li}
is currently a Eng.D. student at Zhejiang University, China.
His research interests include in-network computation, programmable switch application, etc.

\end{IEEEbiography}

\begin{IEEEbiography}[{\includegraphics[width=1in,height=1.25in,clip,keepaspectratio]{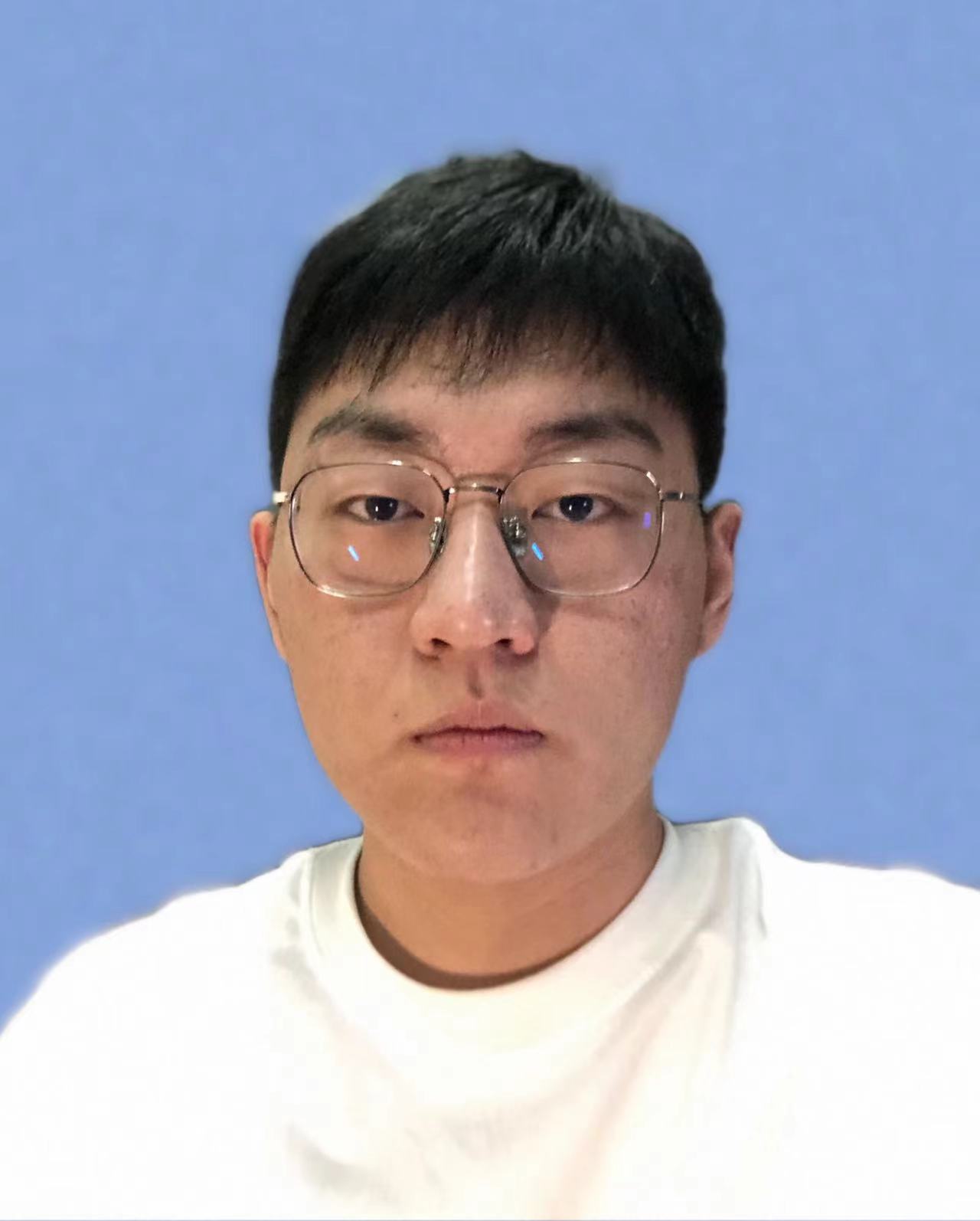}}]{Jie Sun}
is currently a Ph.D. student at Zhejiang University, China. Prior to that, he received his bachelor's degree from Zhejiang University. His research interests include graph neural network, machine learning system, etc.
\end{IEEEbiography}

\begin{IEEEbiography}[{\includegraphics[width=1in,height=1.25in,clip,keepaspectratio]{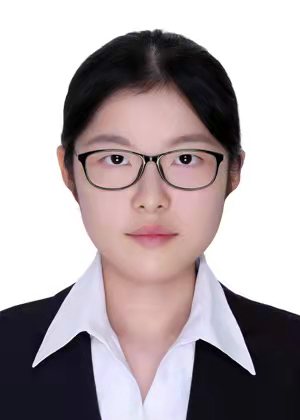}}]{Xueying Zhu}
is currently a Ph.D. student at Zhejiang University, China. Prior to that, she received her bachelor's degree from Zhejiang University. Her research interests include in-network computation, SmartNIC, etc.
\end{IEEEbiography}

\begin{IEEEbiography}[{\includegraphics[width=1in,height=1.25in,clip,keepaspectratio]{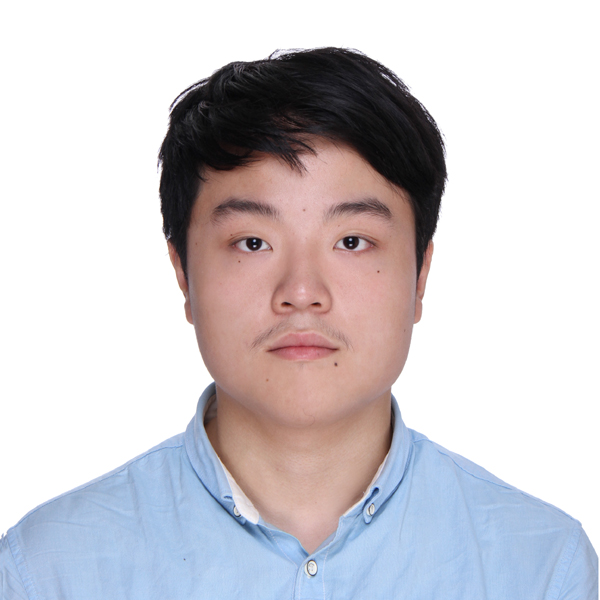}}]{Jie Zhang}
is currently a Ph.D. student at Zhejiang University, China. Prior to that, he received his bachelor's degree from Zhejiang University. His research interests include cloud storage, SmartNIC, etc.
\end{IEEEbiography}

\begin{IEEEbiography}[{\includegraphics[width=1in,height=1.25in,clip,keepaspectratio]{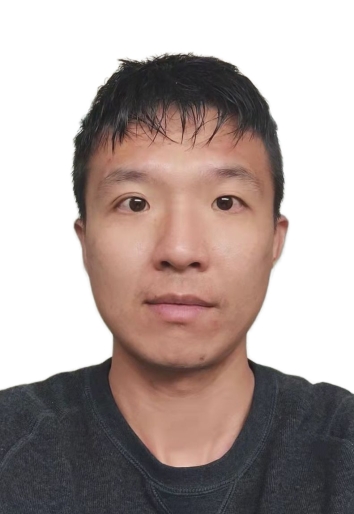}}]{Liang Luo}
Received his Ph.D. degree from University of Washington in 2020. His research focuses on improving distributed training efficiency.
\end{IEEEbiography}

\begin{IEEEbiography}[{\includegraphics[width=1in,height=1.25in,clip,keepaspectratio]{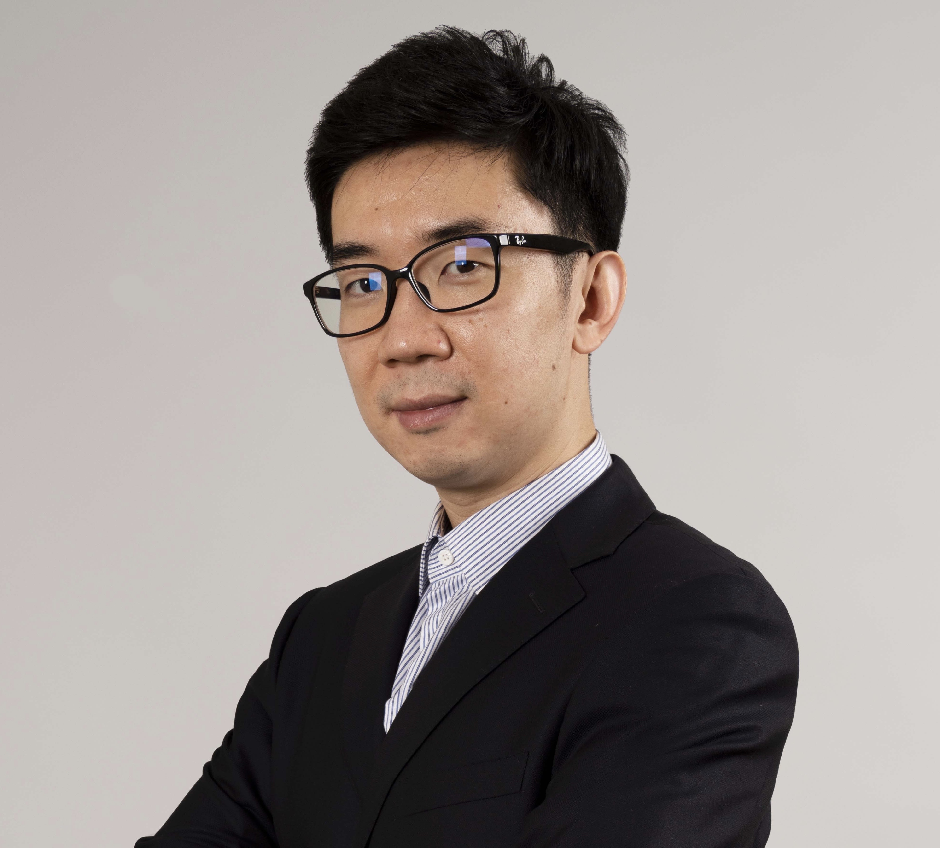}}]{Jialin Li}
received his Ph.D. degree from the University of Washington in 2019. Li is currently an Assistant Professor in the School of Computing at the National University of Singapore. His research interests are in co-designing distributed systems with data center networks, data plane operating systems, and system software for programmable network hardware.
\end{IEEEbiography}

\begin{IEEEbiography}[{\includegraphics[width=1in,height=1.25in,clip,keepaspectratio]{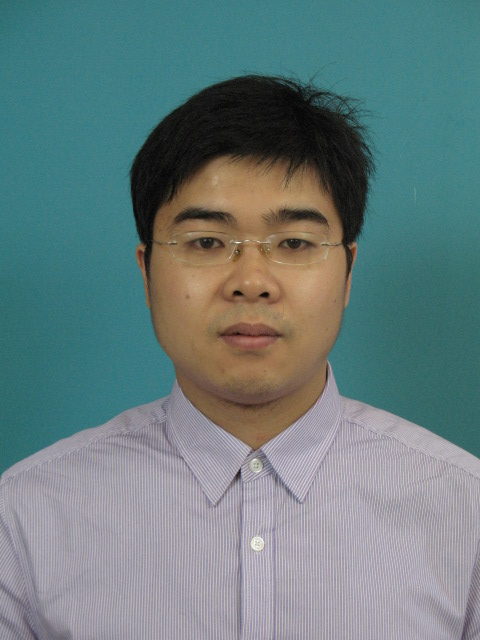}}]{Zeke Wang}
received his Ph.D. degree from Zhejiang University, China in 2011. He is a Research Professor at Collaborative Innovation Center of Artificial Intelligence, Department of Computer Science, Zhejiang University, China. His current research interests mainly focus on building machine learning systems using heterogeneous devices, e.g., SmartNIC and SmartSwitch. 
\end{IEEEbiography}

\end{document}